\documentclass[journal,letterpage,10pt]{IEEEtran}
\usepackage{amsmath}
\usepackage{amsthm}
\usepackage{amsfonts}
\usepackage{amssymb}
\usepackage{makeidx}
\usepackage{cite}
\usepackage{graphicx,bigints}
\usepackage{mathtools}
\usepackage{cases}
\usepackage{color}
\usepackage{hyperref}
\usepackage{bbm}
\usepackage[normalem]{ulem}
\usepackage{braket}
\usepackage{textcomp,gensymb}
\usepackage[normalem]{ulem}
\usepackage{datetime}
\usepackage{enumitem}

\usepackage{epstopdf}
\usepackage{xcolor}

\usepackage[makeroom]{cancel}

\makeatletter
\DeclareFontFamily{U}{tipa}{}
\DeclareFontShape{U}{tipa}{m}{n}{<->tipa10}{}
\newcommand{\arc@char}{{\usefont{U}{tipa}{m}{n}\symbol{62}}}%

\newcommand{\arc}[1]{\mathpalette\arc@arc{#1}}

\newcommand{\arc@arc}[2]{%
	\sbox0{$\m@th#1#2$}%
	\vbox{
		\hbox{\resizebox{\wd0}{\height}{\arc@char}}
		\nointerlineskip
		\box0
	}%
}
\makeatother

\newtheorem{theorem}{Theorem}
\newtheorem{corollary}{Corollary}
\newtheorem{lemma}{Lemma}

\newtheorem{definition}{Definition}
\newtheorem{example}{Example}

\newtheorem{proposition}{Proposition}
\newtheorem{remark}{Remark}

\DeclareMathOperator{\cA}{\mathcal{A}}

\DeclareMathOperator{\cC}{\mathcal{C}}

\DeclareMathOperator{\cO}{\mathcal{O}}

\DeclareMathOperator{\cL}{\mathcal{L}}

\DeclareMathOperator{\SINR}{\textnormal{SINR}}

\DeclareMathOperator{\bR}{\mathbb{R}}

\DeclareMathOperator{\bP}{\mathbf{P}}
\DeclareMathOperator{\ind}{\mathbbm{1}}
\DeclareMathOperator{\bE}{\mathbf{E}}

\newcommand*\diff{\mathop{}\!\mathrm{d}}

\newcommand*\nnb{\nonumber}

\newcommand{\overbar}[1]{\mkern 1.5mu\overline{\mkern-1.5mu#1\mkern-1.5mu}\mkern 1.5mu}

\newcommand\independent{\protect\mathpalette{\protect\independenT}{\perp}}
\def\independenT#1#2{\mathrel{\rlap{$#1#2$}\mkern2mu{#1#2}}}

\definecolor{sandy}{HTML}{E6E2AF}
\definecolor{stone}{HTML}{A7A37E}
\definecolor{beach}{HTML}{EFECCA}
\definecolor{ocean}{HTML}{046380}
\definecolor{diver}{HTML}{002F2F}

\definecolor{Firenze1}{HTML}{468966}
\definecolor{Firenze2}{HTML}{FFF0A5}
\definecolor{Firenze3}{HTML}{FFB03B}
\definecolor{Firenze4}{HTML}{B64926}
\definecolor{Firenze5}{HTML}{8E2800}
\definecolor{mediumpersianblue}{rgb}{0.0, 0.4, 0.65}
\definecolor{hongik}{HTML}{004498}
\definecolor{cobalt}{rgb}{0.0, 0.28, 0.67}
\definecolor{burntorange}{rgb}{0.8, 0.33, 0.0}

\definecolor{ultramarineblue}{rgb}{0.25, 0.4, 0.96}

\title{{Modeling and Analysis of Downlink Communications in a Heterogeneous LEO Satellite Network}}

\author{Chang-Sik Choi,~\IEEEmembership{Member,~IEEE,}
	\IEEEcompsocitemizethanks{\IEEEcompsocthanksitem{Chang-Sik Choi is an Assistant Professor of Dept. of EE, Hongik University, South Korea. (chang-sik.choi@hongik.ac.kr)}}
}
\begin{document}
	\maketitle 
	
	\begin{abstract}
Low Earth Orbit (LEO) satellite networks connect millions of devices on Earth and offer various services, such as data communications, remote sensing, and data harvesting. As the number of services increases, LEO satellite networks will continue to grow, and many LEO satellite network operators will share the same spectrum resources. We aim to examine the coexistence of such a future heterogeneous LEO satellite network by proposing a tractable spatial model and analyzing the basic performance of downlink communications. We model a heterogeneous LEO satellite network as Cox point processes, ensuring satellites are located on various orbits. Then, we analyze two different access technologies for such a heterogeneous satellite network: closed access and open access. For both cases, we derive the coverage probability and prove that the coverage probability of the open access scenario outperforms that of the closed access, and this coverage enhancement applies to all users. By providing essential network performance statistics as key distributional parameters and by presenting the fact that the Cox point process approximates a forthcoming future constellation with slight variation, the developed framework and analysis will serve as a tractable instrument to design, evaluate, and optimize heterogeneous LEO satellite networks with numerous constellations.
	\end{abstract}

\begin{IEEEkeywords}
	Heterogeneous LEO satellite networks, stochastic geometry, Cox point process, user access technology, coverage probability
\end{IEEEkeywords}

	\section{Introduction}
	\subsection{Motivation and Related Work}
\IEEEPARstart{L}{EO} {satellite networks offer global connectivity to devices on Earth without the need for nearby ground infrastructure \cite{6934544,8700141}. Thanks to their large-scale global connectivity, research in LEO satellite networks is receiving a lot of attention both from academia and industry. Currently, only a few standalone LEO satellite network operators exist such as Starlink and OneWeb and they are designed to cover millions of devices on Earth. \cite{okla}. Soon, many companies such as Amazon and Boeing will launch their satellites into orbits \cite{FCCKuiper,FCCBoeing}, resulting in much more complex LEO satellite constellations \cite{8002583,9755278,9970355}.
		
Most satellite communication systems currently use certain microwave bands to serve devices with diverse applications \cite{7060478}. In a system where the number of satellites is small, the impact of interference has only a marginal impact on the operation and the quality of the LEO satellite communications since interference management techniques employed between two or three different parties can ensure the coexistence of independently operated LEO satellite networks \cite{9915631}. Nevertheless, as the number of LEO satellite network operators grows, so as the number of LEO satellites in each constellation, and the inter-constellation or inter-satellite interference will increase sharply, making multi-lateral interference management infeasible. This will undermine the operation of LEO satellite networks and the quality of the downlink communications. This will may get worse as 5G/6G terrestrial networks merge with these non-terrestrial networks \cite{8473417,8571192}. As a result, in the design of LEO satellite networks for the future, the impact of interference generated by various satellite network operators should be taken into account \cite{9915631,10227890,10102783}.
	
The present paper investigates the global impact of such inter-satellite and inter-constellation interference by analyzing a heterogeneous satellite network with multiple constellations with the consideration of two different user access technologies: closed access and open access. Taking a conservative approach, this paper examines an interference-limited environment where satellites of various constellations owned by various operators use the same spectrum resources. In particular, the paper focuses on the geometric features of constellations such as the satellites of different operators are located on their own orbits at different altitudes, as observed in FCC plan \cite{FCCKuiper}.  This study presents a stochastic geometry framework that adequately models this complex spatial distribution of the LEO satellite constellations and evaluates the downlink communications from LEO satellites to ground users on the surface of the Earth. It is worth noting that some studies used stochastic geometry to analyze LEO satellite downlink communications \cite{9313025,9755277}. Specifically, \cite{9079921,9177073,9218989,9497773,9678973,9838263,9841569,9861782,park2023unified,lee2022coverage} employed binomial or Poisson point processes to model the locations of satellites as uniform points on a sphere. Some work leveraged nonhomogeneous point processes to capture the locations of LEO satellites \cite{9681887}. It is worth noting \cite{9841569} conducted research on multi-layer networks, which is also the main subject of the present study.  Nevertheless, these studies did not consider the two essential geometric facts of a heterogeneous LEO satellite network, namely, (i) satellites are exclusively on orbits, and (ii) each constellation has its own various orbits and satellites. Additionally, the interference involves not only satellites from the same constellation but also those from various constellations. To address the first fact, a satellite Cox point process was recently developed \cite{choi2022analytical,choi2023Cox,choi2023dataharvest} where the satellites and orbits are simultaneously produced so that satellites are located exclusively on the orbits, thanks to the Cox structure\cite{chiu2013stochastic,baccelli2010stochastic}. Independently, simple models with orbits were developed by \cite{lee2022coverage,huang2023system}, and yet neither the inter-orbit nor inter-constellation interference was examined in these papers for ease of analysis.
	
To account for the above (i) and (ii), this paper proposes a spatial model that characterizes a heterogeneous LEO satellite network with $K$ constellations distinctive in their numbers of orbits, satellites, and their altitudes. Then, employing tools from stochastic geometry, this paper analyzes the fundamental performance of the downlink communications of such a network by deriving the user association probability, the interference distribution, and the signal to interference-plus-noise ratio (SINR) coverage probability in both closed and open access (Fig. \ref{fig:closeaccess} and \ref{fig:conceptopenaccess-eps-converted-to} ). To the best of the authors' knowledge, this is a pioneering attempt to analyze a heterogeneous LEO satellite network where orbits and satellites from different constellations exhibit different geometric features.

\begin{figure}
	\centering
	\includegraphics[width=.8\linewidth]{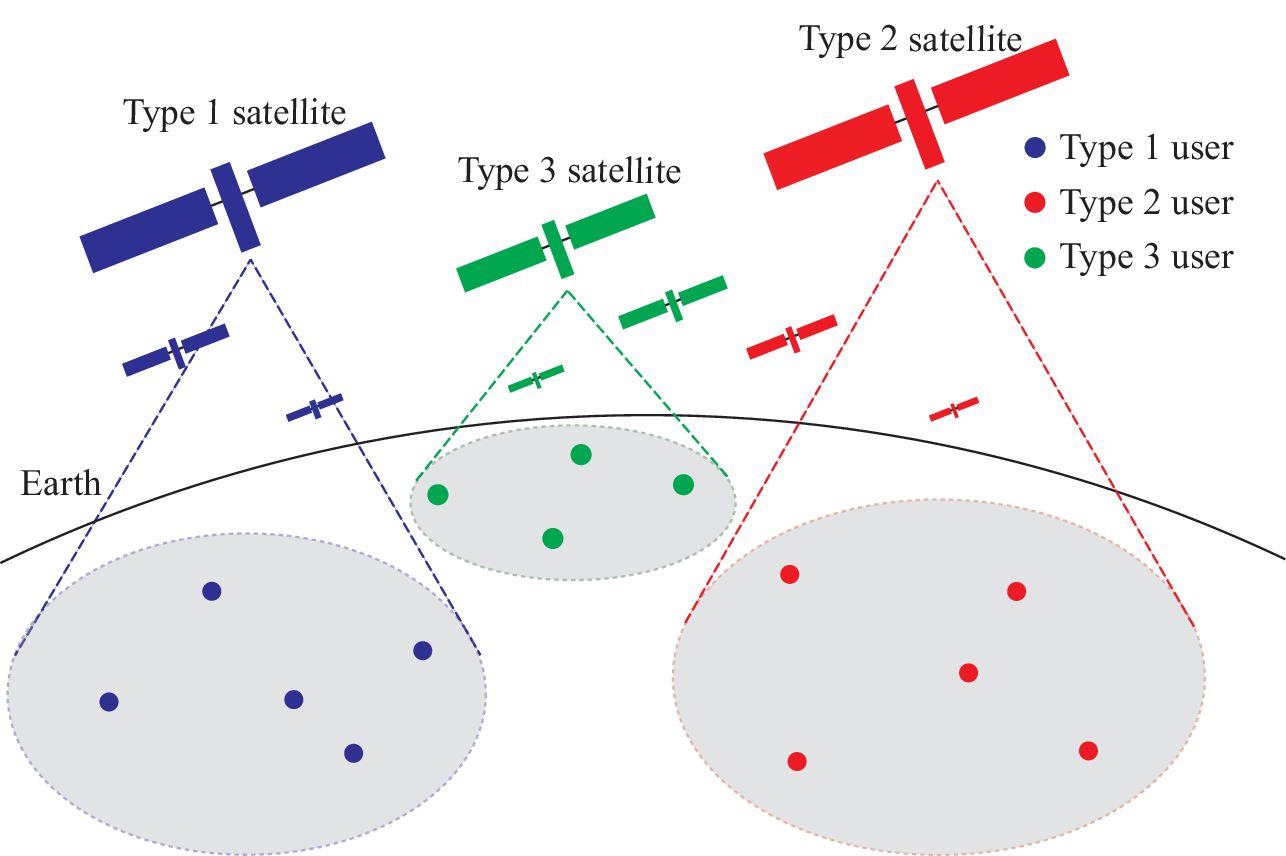}
	\caption{Illustration of the closed access. Users can communicate solely with the LEO satellites of the same constellation type.} 
	\label{fig:closeaccess}
\end{figure}

\begin{figure}
	\centering
	\includegraphics[width=.8\linewidth]{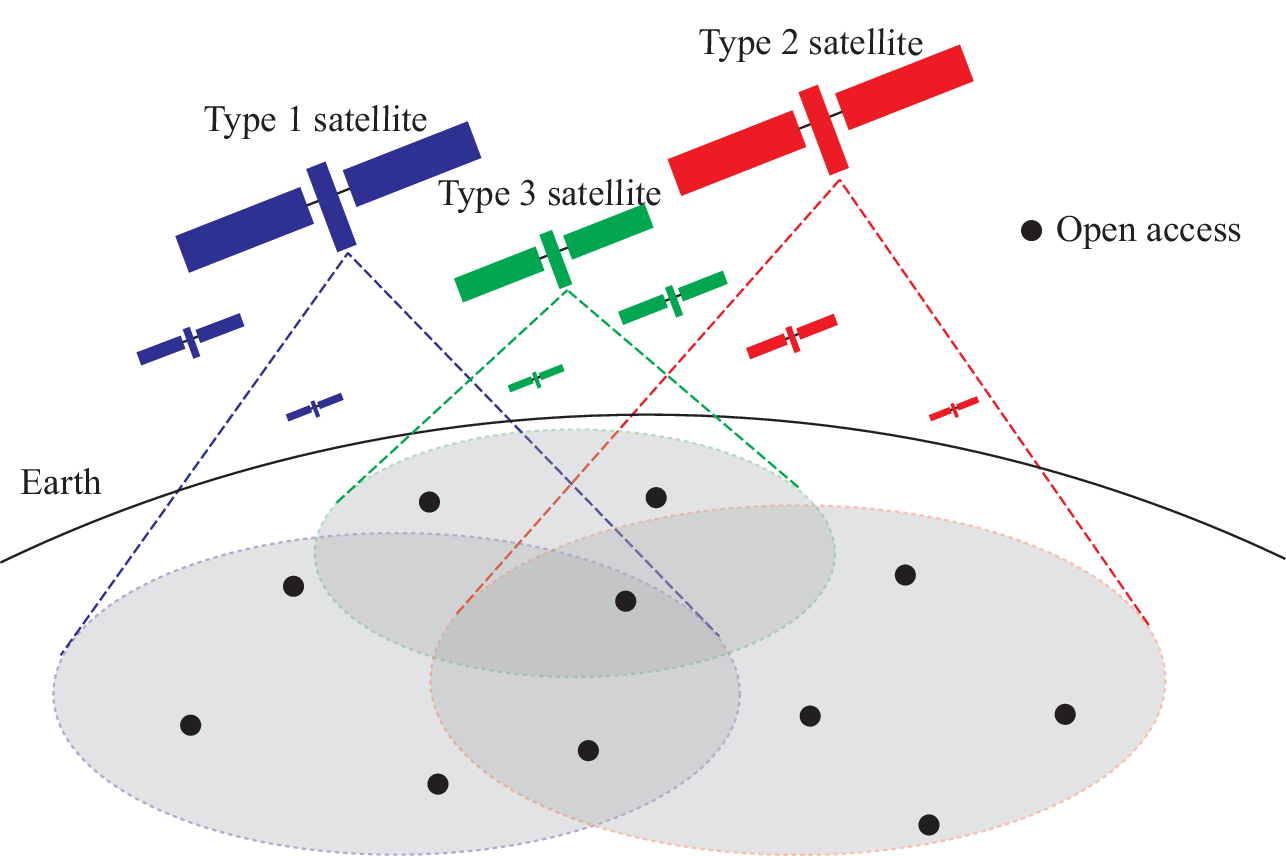}
	\caption{Illustration of the proposed open access scenario. Users can communicate with satellites of any constellation type.}
	\label{fig:conceptopenaccess-eps-converted-to}
\end{figure}

\subsection{Contributions}
The main contributions of this paper are as follows:
 
\uline{Modeling a Future Heterogeneous LEO Satellite Network:} We propose a modeling approach for $K$ independent Low Earth Orbit (LEO) satellite constellations by characterizing them as the superposition of Cox point processes. This modeling approach differs from existing binomial or Poisson satellite models \cite{9079921, 9177073, 9218989, 9497773, 9678973, 9861782}, where all network satellites are uniformly distributed on the sphere at a given altitude. In such models, neither the orbital structure of the LEO satellite network nor the varying altitudes are represented. Our proposed model goes beyond by incorporating the practical geometric aspects of a heterogeneous LEO satellite network, accounting for differences in the number of orbits, the number of satellites, and the altitudes among different constellations. To analyze downlink communications in the network, we first prove the isotropic property of the proposed model. We then provide useful lemmas regarding the arc length, the distance to the Cox-distributed satellite point, and the nearest distance distribution. Later in the study, we demonstrate that, through the moment matching method, the network performance of the Cox point process closely aligns with that of the target constellation, emphasizing the capability of the Cox point process to model existing or upcoming LEO satellite constellations with a marginal error.

\uline{Analysis of interference and coverage probability for closed access:} Using the developed spatial model for a heterogeneous satellite network with $K$ satellite constellations, we first analyze a closed access scenario where the users of a certain type---i.e., the user of a constellation---can communicate only with the LEO satellites of the same type. Leveraging the isotropic property of the proposed satellite point process and its independence from the user point process, we derive the interference seen by a typical user from all $K$ constellations. In the proposed network, the interference comes from not only the LEO satellites of the same type but also the LEO satellites of different types. In particular, for closed access, there are satellites of different types at distances closer than the association satellites from the typical user, leading to an additional high local interference seen by the user. Using the derived Laplace transform of the interference, we evaluate the SINR coverage probability of the typical user and confirm that the present analysis matches the simulation results experimented over various system parameters. For the closed access system, we provide experiments revealing the large-scale behavior of the coverage probability with respect to (w.r.t.) key distributional parameters such as the radii of orbits, the altitudes of satellites, the number of satellites, and the total number of satellite constellations.

\uline{User association and coverage probability for open access}: This paper examines an open access scenario where open access users can communicate with LEO satellites of any type. In particular, we assume network users are associated with their nearest satellites and this eliminates local interfering satellites closer than the association satellite. By leveraging the distribution of the proposed satellite point process and the conditional independence of orbits from different types, this paper derives the probability that a typical user is associated with a certain LEO satellite constellation, as a function of network parameters such as the number of orbits, the number of satellites, and the satellites' altitude. The derived association probability indicates the fraction of connectivity each constellation can offer to users on Earth. Leveraging the association probability, we further derive the Laplace transform of the interference and then the SINR coverage probability of the typical user. By examining both simulation results and analytical formula, we have found that open access sharply improves the coverage of all percentiles users.

	\section{System Model}
This section presents the spatial model for a heterogeneous LEO satellite network with $K$ distinctive satellite constellation types. Then, we discuss user association, signal propagation, and then the performance metrics for the proposed heterogeneous satellite network.
	\subsection{Models for Orbits and Satellites}

\begin{figure}
	\centering
	\includegraphics[width=1.0\linewidth]{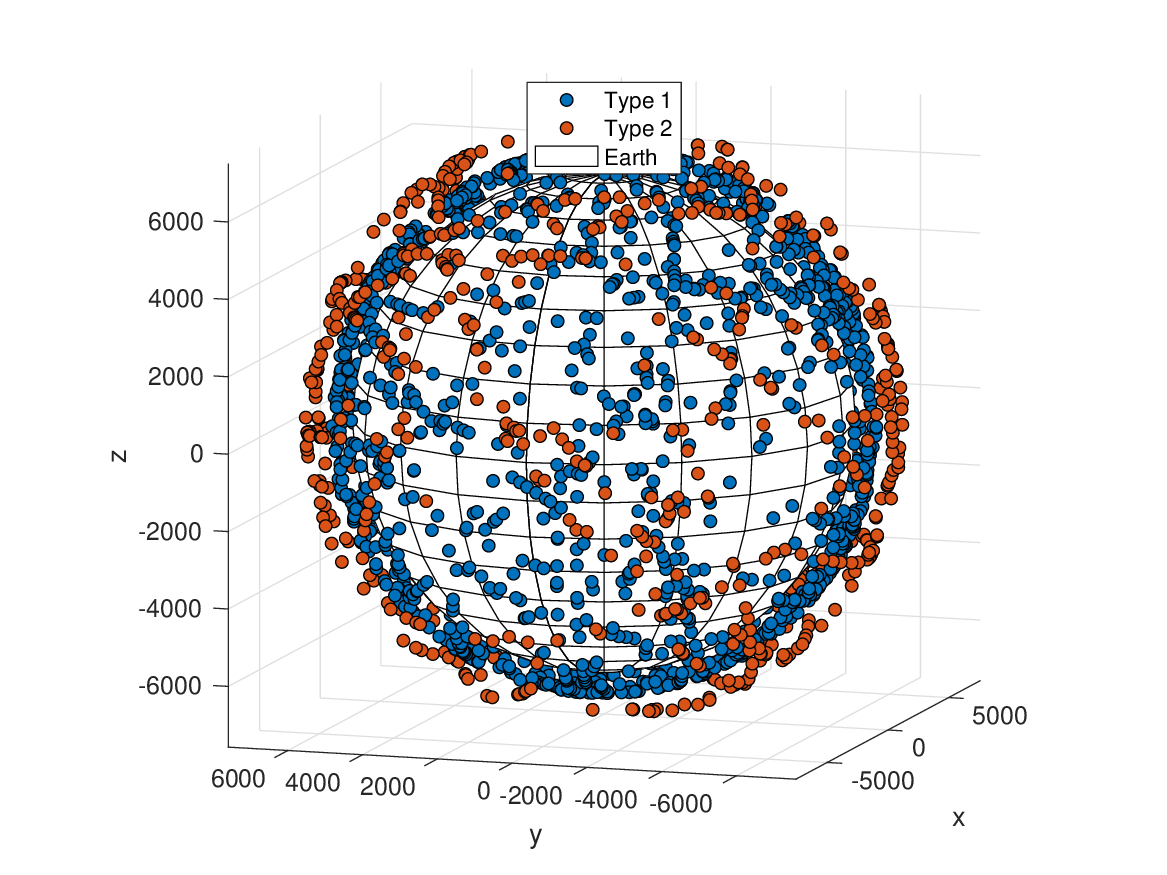}
	\caption{{The Illustration of the network model with $ K=2 $. }}
	\label{fig:modelpicture2}
\end{figure}

\begin{figure}
	\centering
	\includegraphics[width=1.0\linewidth]{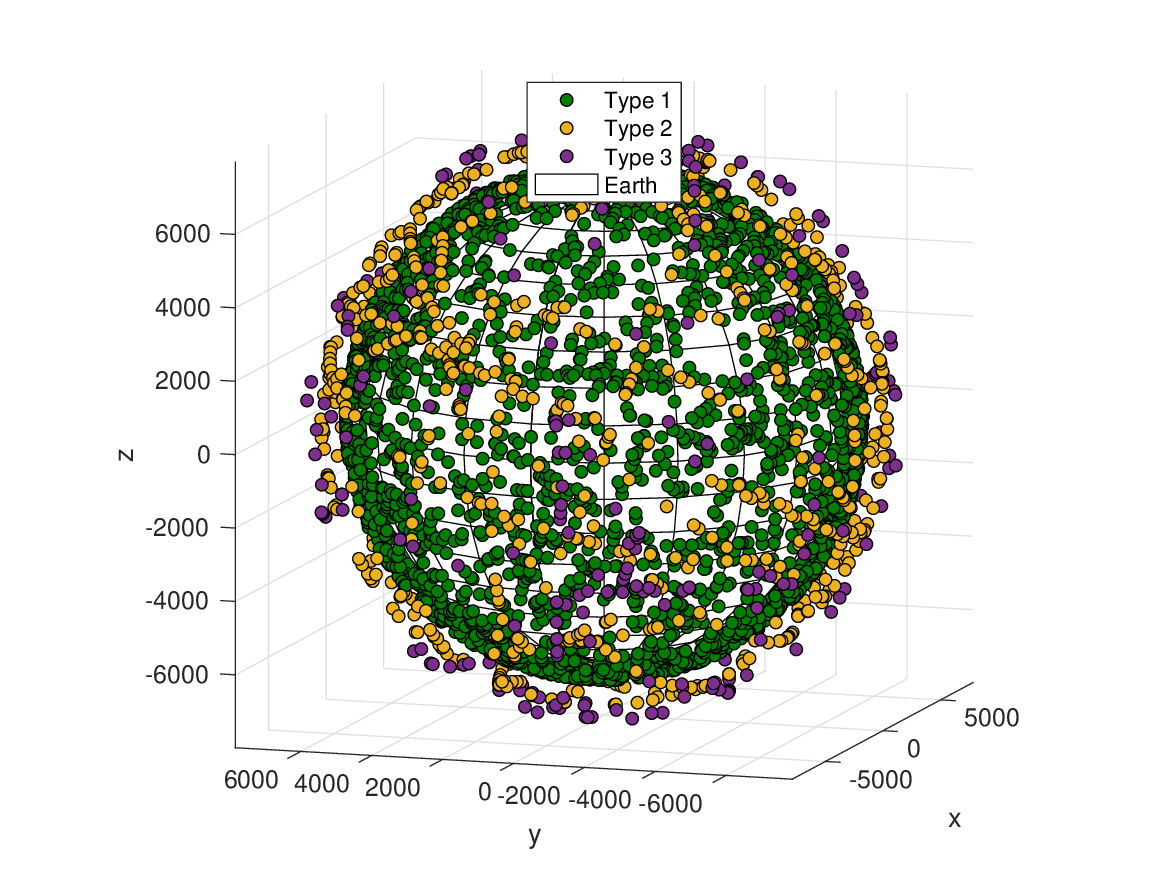}
	\caption{{Illustration of the proposed network with $ K=3 $. }}
	\label{fig:modelpic700036187500241280001812}
\end{figure}

Without loss of generality, we assume that Earth center is $ O=(0,0,0) $ and its radius is $ r_e=6400 $ km. We let the equator to be the $ xy $-plane. The $ x $-axis points to $ 0 $ degree longitude, $ y $-axis points to $ 90 $ degree longitude, and $ z $ axis points to the north pole of the Earth. 
\par To model various orbits of a heterogeneous LEO satellite network, first suppose a cuboid $\cC=[r_a,r_b]\times[0,\pi)\times [0,\pi).$ Then, for the LEO satellite constellation of type $k,$ we use a Poisson point process $\Xi_k$ of density as follows: 
\begin{equation}
 	\lambda_k \nu(\rho)/\pi^2=\frac{\lambda_{k}\delta_{r_k}(\rho)}{2\pi} \diff \rho \sin(\phi)\label{eq:0}
 \end{equation}The Poisson point process for the orbits of type $k$ is given by $ 	 \Xi_{k} = \sum_{i} \delta_{Z_i},$ where its each point ${ (r,\theta,\phi)} $ gives rise to an orbit  $ o(r,\theta,\phi) $ in the Euclidean space $\bR^3$. Specifically, the first coordinate $ r $ is the orbit's radius, $ \theta $ is the orbit's longitude, and $ \phi $ is the orbit's inclination. All orbits are centered at the origin. In Eq. \eqref{eq:0}, $\lambda_k$ is the mean number of orbits in the satellite constellation of type $k$ and $r_k$ is the radii of orbits in the constellation. Note that in general $r_k\neq r_j , \forall k\neq j$ and therefore the developed framework captures the essential geometric fact of a practical heterogeneous satellite network such that independently operated satellite constellations have different altitudes and different populations of orbits and satellites. 
 
 By assuming all $ K $ LEO satellite constellations are independent, we add the Poisson point processes in cuboid to produce 
\begin{equation}
	 \Xi = \sum_{k=1}^K\Xi_k.
\end{equation} This superposition of the Poisson point processes yields the orbit process in the Euclidean space. We have 
\begin{equation}
	\cO = \bigcup_{k=1}^K \cO_k  = \bigcup_{k=1}^K \bigcup_{Z_i\in\Xi_{k}}{o_i},
\end{equation}
where $\cO$ is the orbit process of all constellation types, $ \cO_k $ is the orbit process of type $k$, and $ o_i \equiv o(\rho_{i},\theta_{i},\phi_{i}) $. Note $\Xi,\Xi_k$ are in $\cC $ the cuboid and $\cO,\cO_k$ are in the Euclidean space. 
\par Conditionally on $ \Xi, $ the locations of LEO satellites of type $k$ on  each orbit $ o(r_{k},\theta_{i},\phi_{i}) $ are modeled as a homogeneous Poisson point process $ \psi_i $ of intensity $ \mu_{k}/(2\pi r_{k}) $. The variable $\mu_k$ is the number of satellites of type $k$ per each orbit of the type $k$ satellite constellation. 

Since type $k$ satellites are distributed conditionally on the type $k$ orbit process, the type $k$ satellite point process is a Cox point process \cite{daley2007introduction,chiu2013stochastic}. The proposed satellite Cox point process of all $K$ constellations is now given by the independent summation of the Cox point processes as follows: 
\begin{equation}
	\Psi = \sum_{k=1}^K \Psi_k = \sum_{k=1}^K \sum_{Z_i\in\Xi_k} \psi_{i}.
\end{equation}
Thanks to this unique structure, the proposed model for a heterogeneous LEO satellite network simultaneously creates the orbits and the LEO satellites exclusively on these orbits for each constellation type. 

\begin{figure}
	\centering
	\includegraphics[width=.7\linewidth]{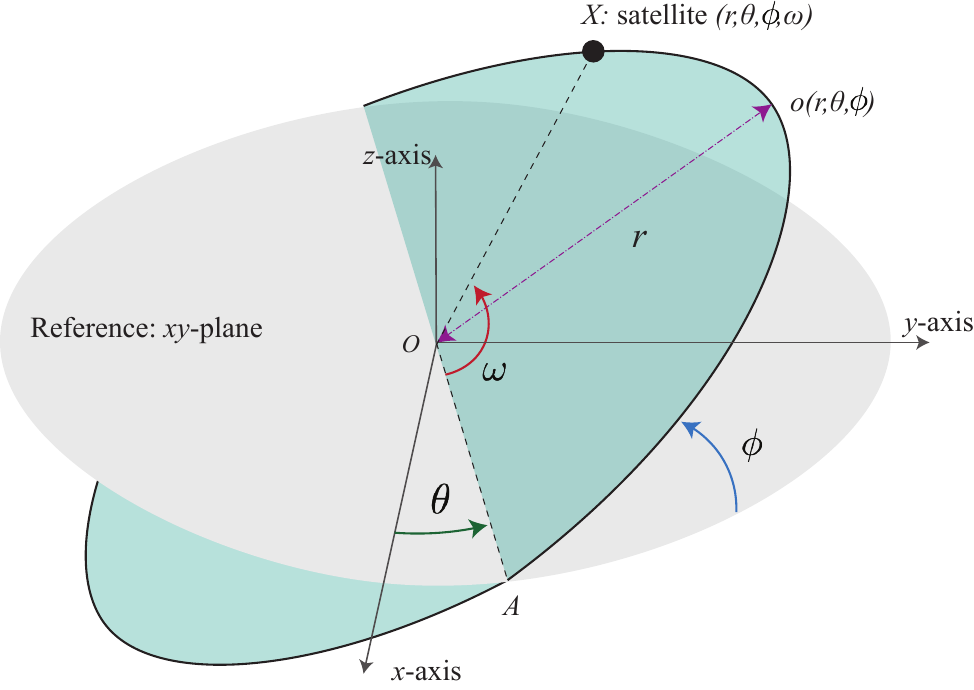}
	\caption{The longitude $ \theta $ is measured from the $ x $-axis to the segment $ \overbar{OA} $. The inclination $ {\phi} $ is measured from the reference plane to the orbital plane. The angle $ \omega $ for satellite $ X $ is measured from $ \overbar{OA} $ to $ \overbar{OX} $ on the orbital plane.}
	\label{fig:shortpaperdefine}
\end{figure}
Figs. \ref{fig:modelpicture2} and \ref{fig:modelpic700036187500241280001812} show the proposed model. In Fig. \ref{fig:modelpicture2}, $K=2$, $\lambda_1=50$, $\mu_{1}=50$, $r_1=6930$, $\lambda_2=20$, $\mu_2=50$, and $ r_2=7600$ km. In Fig. \ref{fig:modelpic700036187500241280001812}, $K=3$, $\lambda_1=50$, $\mu_{1}=50$,  $r_1=6930$ km $\lambda_{2}=20$, $\mu_2=50$,  $r_2=7600$ km and  $\lambda_3=14$, $\mu_3=24$, and $r_3=8000$ km. See Fig. \ref{fig:shortpaperdefine} for the angles $ \theta,\phi, $ and $\omega$ used in this paper. 

\begin{remark}
This study represents a pioneering attempt to model complex orbits involving multiple operators and satellites at different altitudes using stochastic geometry. Previous frameworks have been unable to capture both orbits and satellites simultaneously, resulting in a scarcity of relevant literature. However, to create a model applicable to various deployment scenarios, this paper initially adopts the isotropic orbit process to describe the structure of orbits. These isotropic orbits are essentially generated by a Poisson point process in the cuboid $\cC$. This framework can be extended to reproduce nonuniform altitude orbits by changing the intensity measure of orbit process in a cuboid. For instance, let $ \nu(\diff \rho) = {\diff\rho}/{(r_b-r_a)} $. In this case, the radius of an orbit is a uniformly distributed in the interval $ [r_a,r_b] $.  Similarly, let the intensity measure defined on a subset of the cuboid: $\overbar{\cC}= [r_a,r_b]\times [\theta_m,\theta_M) \times [\phi_m,\phi_M].$ Then, now the longitudes of orbits are distributed in the interval $(\theta_m,\theta_M)$ and the inclinations of orbits are distributed in the interval $(\phi_m,\phi_M).$ In other words, the orbit process is not isotropic.
\end{remark}
 \subsection{User Distribution and Association}\label{S:2-a}
This paper analyzes the following two scenarios: (i) closed access and (ii) open access, analyzed separately under different assumptions. Specifically, Section \ref{S:3} investigates the closed access scenario. Regarding the user distribution and association in the closed access, for $k\in[K]\equiv1,...,K,$  we first assume that type $k$ network users are distributed as an independent Poisson point process of intensity $ \nu_{k} \gg 1$ on Earth and that each user is associated with its nearest LEO satellites of the same type. On the other hand, Section \ref{S:4} investigates the open access scenario. Regarding the user distribution and association in the open access, we assume that open access users are distributed as an independent Poisson point process of density $ \nu\gg 1 $ on Earth and that each user is associated with its nearest LEO satellite of any type $k\in [K]$. In contrast to the closed access, the open access assumes network users to be associated with their nearest LEO satellites regardless of constellations. To facilitate analysis, we assume that the distributions of users and LEO satellites are independent for both scenarios.
\begin{table}
	\centering
	\caption{User Distribution and Association}
	\begin{tabular}{|l|l|l|}
		\hline 
		Scenario & Users distribution & User association \\\hline
		Closed access &  $K$ independent Poisson & Same type nearest \\\hline 
		Open access &  independent Poisson & Any nearest\\\hline 
	\end{tabular}
\end{table}

\subsection{Propagation Model}
The signals from LEO satellites to network users attenuate. To emphasize the role of network geometric variables and to facilitate analysis, this paper assumes a simple propagation model as in  \cite{6184256,9218989,9497773,9678973}. The received signal power at distance $d (d>1)$ from LEO satellite transmitter of type $k$ is given by 
	\begin{equation}\label{rxpower}
		p_{k} g_k H d^{-\alpha},
	\end{equation}
	where $ p_{k} $ is the transmit power of type $k$ satellite, $ g_{k} $ is the aggregate array gain from the transmit and receive antennas associated with the constellation of type $k$, $ H $ is small-scale fading, and $ \alpha $ is the path loss exponent. We assume that for a desirable signal $g_k\neq 1$; for interference,  $g_k=1.$ 
		 		

	 		\par For small-scale fading, we consider Nakagami-$ m $ fading incorporating both LOS and NLOS propagation scenarios, as considered in  \cite{9861782,lee2022coverage}. The CCDF of random variable $ {H} $ is 
	 		\begin{align}
	 			\bP(H>x) &= 
	 				e^{-mx}\sum\limits_{k=0}^{m-1} \frac{(m x)^k}{k!} &\forall x\geq 0,
	 			\label{CDF of H}
	 		\end{align}
 	where $ m\geq 1 $ and $ \Gamma(\cdot) $ is the Gamma function. If $ m=1$, Nakagami-$ m $ fading becomes Rayleigh fading and $ H $ follows an exponential random variable with mean $ 1 $. 
\begin{remark}
This work conducts an initial investigation of the coverage probability in a heterogeneous LEO satellite network, excluding non-geometric aspects such as Gaussian channel gain or shadowed-Ricean small-scale fading \cite{9678973}. It is important to note that, using the proposed spatial model, the impact of a Gaussian beam pattern can still be examined by assuming additional antenna gain of the interfering satellites close to the typical user. However, this introduces a second-order spatial correlation between satellites, which complicates the derivation of the coverage probability. Similarly, network performance analysis can incorporate shadowed-Ricean small-scale fading by replacing Eq. \eqref{CDF of H} with the appropriate distribution function for shadowed-Ricean fading. A more detailed analysis is left for future work. 
 		\end{remark}
	\subsection{Performance Metrics}
	\begin{itemize}
		\item No-satellite: Some users on Earth may not discover any visible satellite for a given time. This paper defines the no-satellite probability as the probability that the typical user cannot  find any visible satellite of the same type (for the closed access) or of any type (for the open access). 
		\item Closed access coverage probability: The closed access coverage probability of the network user at $ y $ of type $ k $ is 
		\begin{align*}
			\bP(\SINR_{\textnormal{cl},k}>\tau)=& \bP\left(\frac{p_{{k}} g_kH \|X_y-y\|^{-\alpha}}{\sigma^2 + I_{y;c} }> \tau \right),
		\end{align*}
		where $ y $ is the location of the network user, $ X_y $ is the association LEO satellite location, $ \sigma^2 $ is the noise power, and $ {I_{y;c}} $ is the interference seen by the user $ y $  in the proposed closed access LEO satellite network. The interference can be separated into two components: inter-satellite interference generated by the same type of LEO satellites at distances further than $ X_y $ from $ y $ and the inter-constellation interference incurred by LEO satellites of different types. 
		\item Open access coverage probability: In the open access scenario, users can communicate with their nearest satellites, regardless of their types. Let $ t(X) $ be the type of the satellite $ X. $ Then, the open access coverage probability of the network user at $ y $ is  
		\begin{align*}
			&\bP(\SINR_{\text{op}}>\tau)=\bP\left(\frac{p_{{t(X_y)}} g_{t}H \|X_y-y\|^{-\alpha}}{\sigma^2 + I_{y;o} }> \tau \right),
		\end{align*}
		where $ X_y $ is the association LEO satellite location, $ p_{t(X_y)} $ is its transmit power, $ g_{t} $ is the aggregate array gain from the transmit and receive antennas,  and $ I_{y;o} $ is the open access interference at $y$. Note the SINR evaluation in the open access scenario differs from that of the closed access scenario because of the fact that network users are now able to communicate with nearest LEO satellites of any type.
	\end{itemize}

 \begin{remark}
This work aims to investigate the impact of geometric parameters on the performance of downlink communications from LEO satellites to ground users, specifically focusing on the physical layer aspects of these communications in heterogeneous LEO satellite networks. While there are other important components in a heterogeneous satellite network at the access control layer or transport layer, such as inter-satellite links, satellite gateway deployment, uplink capability, and traffic offloading, they fall outside the scope of this study, and their influence on the system is not addressed in this paper. In essence, this paper assumes that users can receive downlink signals from LEO satellites, and these satellites have data to deliver to those users. A similar approach has been taken in the analysis of downlink communications in various network architectures, where the primary focus is on major downlink physical layer components, such as association, interference, and the SINR coverage probability \cite{6171996}.
 \end{remark}

\begin{table}
	\centering
	\caption{Network Variables}\label{Table:1}
	\begin{tabular}{|c|l|}
		\hline
		Variable & Description \\
				\hline
		$K$& Number of heterogeneous satelite constellations\\ 
		\hline
		$\Xi_k$& Poisson process for creating type $k$ orbit\\ 
		\hline
		$\Xi $& Poisson process for creating all orbit\\
		\hline
		$\cO_k$& Type $k$ orbit process  \\
		\hline
		$\cO $& Orbit process of all type  \\
		\hline
				$o_i$& An orbit generated by a point $Z_i=(r_{i},\theta_{i},\phi_{i})$ \\\hline
						$(r_i,\theta_i,\phi_i)$& An orbit's radius, longitude, and inclination \\\hline
		$\Psi_k$& Cox point process for type $k$ satellites\\
		\hline
		$\Psi $& Cox point process of all satellite types\\
\hline
$\lambda_k$ & Mean number of type $k$ orbits  \\ 
\hline
$\mu_k$ & Mean number satellites on each type $k$ orbit\\ 
\hline
$\lambda_k\mu_k$ & Mean number type $k$ satellites \\ 
\hline
$H$ & Fading random variable \\\hline
$\alpha$ & path loss exponent \\\hline
$g_k$ & Aggregate antenna gain for type $k$ constellation \\\hline
	\end{tabular}
\end{table}
\section{Preliminaries: Cox Point Process for Satellites}
This section presents essential statistical properties of the proposed Cox point process for multi-altitude satellite constellations.
\begin{lemma}
	Consider the type $k$ satellite Cox point process of the orbital density $\lambda_k$ and of the satellite density $\mu_k$. The average number of all types of satellites is $\sum_{k=1}^K \lambda_k\mu_k$.
\end{lemma}
\begin{IEEEproof}
The average number of LEO satellites of type $k$ constellation is given by $\lambda_l\mu_{k}$. Then, 	since $\{\Psi_{k}\}_{k=1,...,K}$ are independent Cox point processes and $\sum_{k=1}^{K}\Psi_k$, the average number of LEO satellites of all types is  $\sum_{k=1}^K \lambda_k\mu_k$.
\end{IEEEproof}

\begin{lemma}\label{Lemma:2}
	The proposed orbit process $ \Xi $ and the satellite Cox point process $ \Psi $ are isotropic, namely rotation invariant.
\end{lemma}
\begin{IEEEproof} 
We know that a Cox point process of type $k$ LEO satellites, denoted by $\Psi_k$ is isotropic \cite{choi2023Cox}. Since $ \Psi = \sum_{k=1}^K \Psi_k$ and $\Psi_k\independent \Psi_j,k\neq j$, we conclude that $\cO $ and $ \Psi $ proposed in this paper are isotropic. 
\end{IEEEproof}
Because of this isotropic property and the fact that network users are independent of the satellite Cox point process, we feature a typical user at $U=(0,0,r_e)$ and derive the network performance seen by the typical user. 

More importantly, since $\cO$ and $\Psi$ are rotation invariant, the network performance seen by this typical user represents the network performance of all the users, spatially averaged across all the network. 
\begin{lemma}\label{lemma:2}
	Consider an orbit $ o(r_k, \theta_{i},\phi_{i}) $ of type $k$ and its satellite with angle $ \omega_j$. The distance from the typical user to the satellite $ X $ is given by  
	\begin{align*}
		f_{k,\phi_i}(\omega_j)
		&=\sqrt{r_k^2-2r_kr_e\sin(\omega_j)\sin(\phi_i)+r_e^2}.
	\end{align*}
\end{lemma} 
\begin{IEEEproof}
	See \cite{choi2023Cox}.
\end{IEEEproof}

Let $ \mathbb{S}_{r} $ be the sphere of radius $ r$ centered at the origin.

\begin{definition}
	The spherical cap $ C(r,d) $ is defined by 
	\begin{align*}
		\{(x,y,z)\in\mathbb{S}_{r} \text{ s.t. } \|(x,y,z)-U\|\leq d \}.
	\end{align*}
In other words, $C(r,d)$ is the set of  points on $\mathbb{S}_{r}$ at distances less than $d$. 
\end{definition}
The following lemma evaluates the arc length of the intersection of an orbit and a spherical cap. We use it to compute the void probability of the satellite Poisson point process conditionally on orbit.

\begin{lemma}\label{Lemma:3}
	For $ r - r_e \leq d\leq \sqrt{r^2-r_e^2},  $ the length of the arc given by the intersection of the spherical cap $ C (r,d) $ and the orbit $ o(r,\theta,\phi) $ is given by 
	\begin{equation}
 2r\arcsin\left(\sqrt{1-\cos^2(\xi)\csc^2(\phi)}\right),
\end{equation}
where $ \cos(\xi)  = {(r^2+r_e^2-d^2)}/{(2r_e r)}$ and $ \xi $ is the angle to the rim of the spherical cap $C(r,d)$, measured from the positive $z$-axis. 
\end{lemma}
\begin{IEEEproof}
	See \cite{choi2023Cox}.
\end{IEEEproof}
\begin{lemma}\label{Lemma:5}
	Let $D_k$ be the distance from the typical user $ U $ to its closest visible satellite of type $ k $. When there is no visible satellite, we let $D_k =\infty.$ 
	
	Then, its CCDF $ \bP(D_k>v) $ is $1$ for $0<v<r_k-r_e$ and  for $r_k-r_e\leq v\leq \sqrt{r_k^2-r_e^2}$, 
	\begin{align}
		\exp\left(-{\lambda_k}\int_0^{\xi_k}\right.\!\!\!
		\left.\left(1-e^{-\frac{\mu_{k}}{\pi}\sin^{-1}\left(\sqrt{1-\frac{\cos^2(\xi_k)}{\cos^2(\varphi)}}\right)}\right)\! \cos(\varphi)\diff \varphi \right),\nnb
	\end{align}
where $ \cos(\xi_k) = (r_k^2+r_e^2-v^2)/(2r_kr_e)$, and lastly for $\sqrt{r_k^2-r_e^2}<v, $ it is given by 
	\begin{align}
	&\exp\left(-{\lambda_{k}}\!\int_0^{{\overbar{\varphi}_{k}}}\right.\nnb\\
	&\hspace{17mm}\left.\left(1-e^{-\frac{\mu_{k}}{\pi} \sin^{-1}(\sqrt{1-r_e^2\cos^2(\varphi)/r_{k}^2})}\right)\cos(\varphi)\diff \varphi\right),\nnb
\end{align}
where  $ \overbar{\varphi}_k=\arccos(r_e/r_k)$. 
\end{lemma}
\begin{IEEEproof}
	See \cite{choi2023Cox}.
\end{IEEEproof}

\section{Closed Access}\label{S:3}
We first derive the closed access no-satellite probability of the proposed network and then analyze the SINR coverage probability of the typical user. 
\subsection{Closed Access No-Satellite Probability}
\begin{proposition}\label{prop:1}
	The closed access no-satellite probability of type $k$ constellation, denoted by $\bP(\textnormal{no-satellite}_{k})$, is 
	\begin{align}
		&\exp\left(-{\lambda_{k}}\!\int_0^{{\overbar{\varphi}_{k}}}\right.\nnb\\
		&\hspace{17mm}\left.\left(1-e^{-\frac{\mu_{k}}{\pi} \sin^{-1}(\sqrt{1-r_e^2\cos^2(\varphi)/r_{k}^2})}\right)\cos(\varphi)\diff \varphi\right),\nnb
	\end{align}
	where $ {\overbar{\varphi}_{k}}=\arccos(r_e/r_k)$. Here, the variables $\lambda_k,\mu_k$, and $r_k$ are the average number of orbits, the average number of satellites per orbit, and the altitude of orbits, respectively, for the type $k$ constellation. 
\end{proposition}
\begin{IEEEproof}
	In the closed access heterogeneous LEO satellite network, the typical user of type $k$ is associated only with the same type satellite. Therefore, the no-satellite probability of type $k$ constellation is the same as the probability that there is no visible satellite of type $k,$ namely $\bP(D_k = \infty).$ In addition, since $\Psi_k \independent \Psi_i, i\neq k,$ we use Lemma \ref{Lemma:5} to have the result. 
\end{IEEEproof}

\begin{figure}
	\centering
	\includegraphics[width=1.0\linewidth]{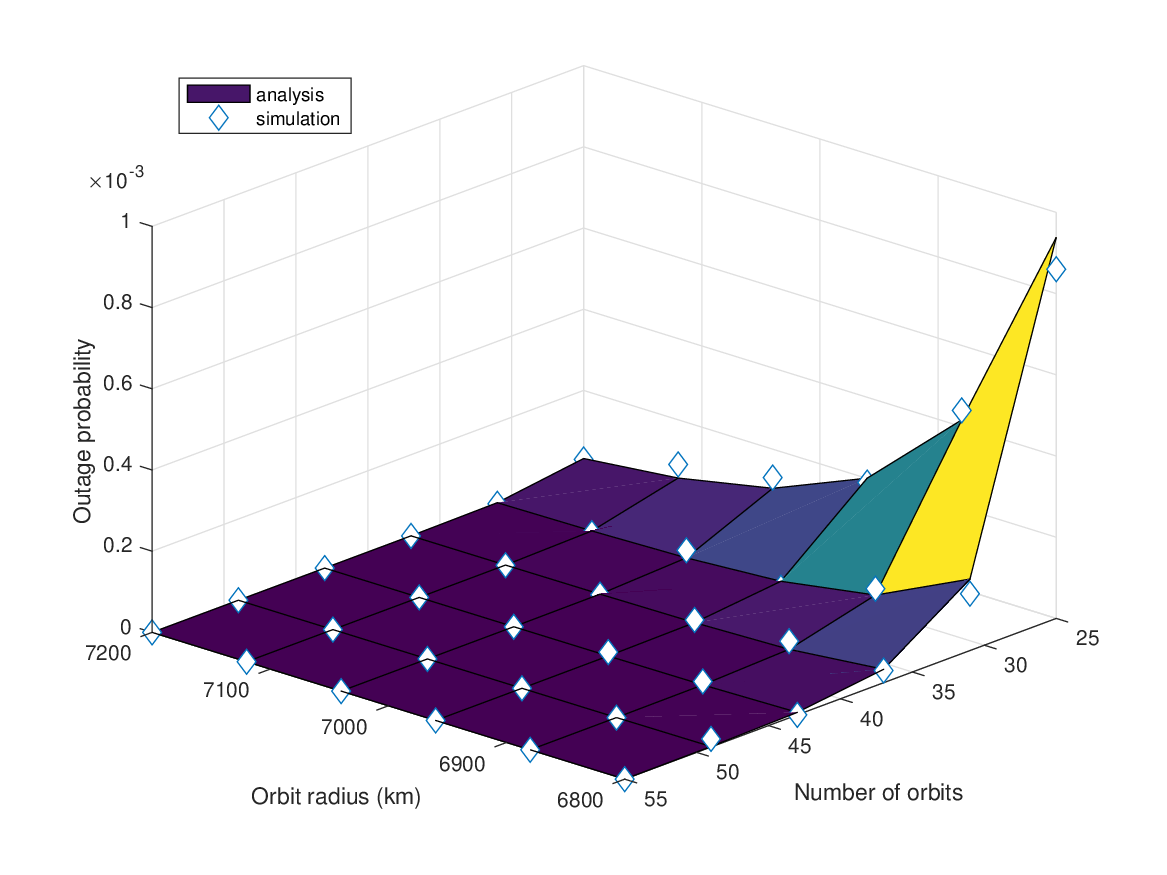}
	\caption{The no-satellite probability of type $k=1$ user. }
	\label{fig:propositionone}
\end{figure}

The aforementioned proposition applies to any $k\in[K]$. Figure \ref{fig:propositionone} illustrates the no-satellite probability obtained by both simulation and formula. The simulation results validate the accuracy of the derived formula. We use $\mu_k=22$. 

When the number of satellites is not high, e.g., $\lambda_k=25 $, $\mu_{k}=22 $, (total $400$ satellites on average) and the satellite altitude is low, e.g., $400$ km, the no-satellite probability is about $0.001$. This means one out of $1000 $ users may experience an no-satellite. Nevertheless, as the radius of orbits is sufficiently large (e.g., greater than $7000$ km) and that the mean number of orbits is greater than $40$, the no-satellite probability becomes very low less than $10^{-5}$. 

It is worth noting that the no-satellite probability of the typical user of type $k$ is independent of the distribution of other constellations. This independence arises because the typical user of type $k$ is allowed to communicate exclusively with the same type of satellite.

\subsection{Closed Access Coverage Probability}
In following, to facilitate analysis, we compute the SINR coverage probability of the typical user assuming that $\lambda$ and $\mu$ are sufficiently large so that no-satellite probability is very small.


\begin{figure*}
	\begin{align}
		&\int_{r_k-r_e}^{\sqrt{r_k^2-r_e^2}}\frac{\lambda_k \mu_k z }{ \pi r_kr_e}\left(e^{-\frac{\sigma^2 \tau z^\alpha}{p_kg_k}}\prod_{l=1,l\neq k}^{K}\cL_{I_{\text{cl},l}}\left(\frac{\tau z^\alpha}{p_kg_k}\right)\right)\nnb\\
		&\hspace{15mm}\exp\left(-{\lambda_{k}}\int_{0}^{\xi_{k}}\left(1- e^{-\frac{\mu_k}{\pi}\arcsin(\sqrt{1-\cos(\xi_k)^2\sec(\varphi)^2})-\frac{\mu_{k}}{\pi}\int_{\omega_{k,\varphi,1}}^{\omega_{k,\varphi,2}}\left(1-\cL_H(\tau z^\alpha /(g_k\overbar{f}_{k,\varphi}^{\alpha}(\omega)))\right)\diff \omega}\right)\cos(\varphi)\diff \varphi\right)\nnb\\
		&\hspace{15mm}\exp\left(-{\lambda_{k}}\int_{\xi_{k}}^{\bar{\varphi}_k}\left(1- e^{-\frac{\mu_{k}}{\pi}\int_{0}^{\omega_{k,\varphi,2}}\left(1-\cL_H(\tau z^\alpha /(g_k\overbar{f}_{k,\varphi}^{\alpha}(\omega)))\right)\diff \omega}\right)\cos(\varphi)\diff \varphi\right)\nnb\\ 
		&\hspace{15mm}\left(\int_{0}^{{\xi_k}}\frac{e^{-\frac{\mu_k}{\pi}\arcsin(\sqrt{1-\cos(\xi_k)^2\sec(v)^2})-\frac{\mu_{k}}{\pi}\int_{\omega_{k,v,1}}^{\omega_{k,v,2}}\left(1-\cL_H(\tau z^\alpha /(g_k\overbar{f}_{k,v}^{\alpha}(\omega)))\right)\diff \omega}}{\sqrt{1-\cos^2(\xi_k)\sec^2(v)}}\diff v\right)  \diff z \label{eq:theorem:1}.
	\end{align}
	\rule{\linewidth}{0.1mm}
\end{figure*}

\begin{theorem}\label{Theorem:1}
	When $ m=1 $, the SINR coverage probability of the type $k$ typical user is Eq. \eqref{eq:theorem:1}. 
\end{theorem}
\begin{IEEEproof}
	The closed access coverage probability of the type $k$ typical user is given by 
	\begin{align}
	&\bP(\SINR_{\textnormal{cl},k}\geq\tau)\nnb\\
	&= \bP\left(\frac{p_{{k}} g_kH }{ (\sigma^2+I_{\text{cl}}) \|X_k^\star-U\|^{\alpha}}\geq \tau \right)\nnb\\
		&=\bE\left[\bE\left[\bE\left[\bP\left(\left.H\geq\frac{\tau (\sigma^2+I_{\text{cl}})z^{\alpha}}{p_kg_k}\right| \Xi_k,Z_\star , d\right) \right]\right]\right]\nnb\\
	&=\bE\left[\bE\left[\bE\left[ e^{-\frac{\tau\sigma^2d^\alpha}{p_kg_k}}\cL_{I_{\text{cl}}}\left( \frac{\tau z^\alpha}{p_kg_k}\right)\right]\right]\right]\label{eq:SINR},
\end{align}
where $ X_k^\star $ is the association satellite of type $ k $ and  $ d= \|X_k^\star-U\|.$ Let $ \cL_{I_{\text{cl}}}\left( {\tau z^\alpha}/{(p_kg_k)}\right) $ be the Laplace transform of the interference evaluated at ${\tau z^\alpha}/{(p_kg_k)},$ conditionally on $ \Xi_{k},Z_\star$, and $ d. $ 	

The interference seen by the type $k$  typical user is given by 
\begin{align}
	I_{\text{cl}}
	=I_{\text{cl},k}^{X_k^\star}+I_{\text{cl},[K]\setminus k}
	&=\sum_{X_j\in \bar{\Psi}_{{k}}\setminus X_{U}} p_{{k}} H\|X_j- U \|^{-\alpha}\nnb\\
	&\hspace{4mm}+\sum_{l=1,l \neq {k}}^{K}\sum_{X_j\in {\bar{\Psi}_l}} p_l H\|X_j- U \|^{-\alpha},\nnb
\end{align}
where  $ I_{\text{cl},k}^{X_k^\star} $ is the interference from $ \bar{\Psi}_k\setminus X_k^\star $ with $ \bar{\Psi}_l $ be the set of visible satellites of type $ l $ and $ I_{\text{cl},[K]\setminus k} $ is the interference from $ \{\bar{\Psi}_l\}_{l\in[K]\setminus k} $. 
Since the satellite Cox point process of different types are independent, we can write the Laplace transform of the interference as follows: 
\begin{align}
	\cL_{I_{\text{cl}}}\left( s \right) 
	&=\cL_{I_{\text{cl}, k}^{X_U}}(s)\prod_{l=1,l\neq k}^K\cL_{I_{\text{cl},l}}(s).
	\label{eq:14}
\end{align}

In Eq. \eqref{eq:14}, the Laplace transform of the interference from satellites of type $l$ is given by 
\begin{align}
	&\cL_{I_{\text{cl},l}}(s) \nnb\\
	&=\bE\left[\prod_{Z_i\in{\Xi}_{l}}^{|\phi_i-\pi/2|<\bar{\varphi}_l}\prod_{X_j\in{\bar{\psi}}_{l,i}} e^{-s p_l  H \|X_j - U\|^{-\alpha}}\right]\nnb\\
	&=\bE\left[\prod_{Z_i\in{\Xi}_{l}}^{|\phi_i-\pi/2|<\bar{\varphi}_l}\!e^{-\frac{\mu_{l}}{2\pi}\int_{\pi/2-\omega_{l,\phi_i,2}}^{\pi/2+\omega_{l,\phi_i,2}}\left(1-\cL_H(sp_l/({f}_{l,\phi_i}^{\alpha}(\omega)))\right)\diff \omega}\right]\nnb\\	&=\bE\left[\prod_{Z_i\in{\Xi}_{l}}^{|\phi_i-\pi/2|<\bar{\varphi}_l}\!e^{-\frac{\mu_{l}}{\pi}\int_{0}^{\omega_{l,\phi_i,2}}\left(1-\cL_H(sp_l/(\widetilde{f}_{l,\phi_i}^{\alpha}(\omega)))\right)\diff \omega}\right]\label{167},
\end{align}
where $ \bar{\varphi}_l= \arccos(r_e/r_l) $  the maximum of the complementary inclination angles of orbits that are visible to the typical user. Above, $ \bar{\psi}_{l,i} $ is the set of visible satellites of type $ l$ on the orbit $ o_i $. To get Eq. \eqref{167} we use the probability generating functional of a Poisson point process of intensity $\mu_l$\cite{daley2007introduction,baccelli2010stochastic,chiu2013stochastic}.  Specifically, leveraging Lemma \ref{Lemma:3}, we get the angles of the visible interfering satellites as follows: $ \{\pi/2-\omega_{l,\phi_i,2}<\omega<\pi/2+\omega_{l,\phi_i,2}\}$ where  
\begin{align}
		{f}_{l,\phi_i}(\omega_j)&=\sqrt{r_l^2-2r_lr_e\sin(\omega_j)\sin(\phi_i)+r_e^2},\\
	\omega_{l,\phi_i,2}&=\arcsin(\sqrt{1-(r_e/r_l)^2\csc^2(\phi_i)}),
\end{align}  where  we use Lemma \ref{lemma:2}. Then, by employing  the change of variables, we obtain Eq. \eqref{167} where 
\begin{align*}
	\tilde{f}_{l,\phi_i}(\omega_j)=\sqrt{r_l^2-2r_lr_e\cos(\omega_j)\sin(\phi_i)+r_e^2}.
\end{align*}  

Finally, by applying the change of variable  $\phi = \pi/2-\varphi$ and then the probability generating functional of the Poisson point process $ \Xi_l $, we arrive at 
\begin{align}
	&\cL_{I_{\text{cl},l}}(s)\nnb\\ 
	&=\exp\left(\!-{\lambda_l}\!\!\int_0^{\bar{\varphi}_l}\!\!1-e^{-\frac{\mu_{l}}{\pi}\int_{0}^{\omega_{l,\varphi,2}}1-\cL_H\left(\frac{sp_l}{\overbar{f}_{l,\varphi}^\alpha(\omega)}\right)\diff \omega}\diff \varphi\right).\label{eq:18}
\end{align}
where we have 
\begin{align}
	\overbar{f}_{l,\varphi}(\omega)&=\sqrt{r_l^2-2r_lr_e\cos(\omega)\cos(\varphi)+r_e^2},\\
		\omega_{l,\varphi,2}&=\arcsin(\sqrt{1-(r_e/r_l)^2\sec^2(\varphi)}).\label{eq:omega1}
\end{align} 

This expression precisely account for the geometric characteristic of the closed access scenario such that there are interfering satellites of different types that are closer than the association satellite.

On the other hand, to get $\cL_{I_{\text{cl}, k}^{X_U}}(s)$ of Eq. \eqref{eq:14}, we use the fact that association satellite $X_U\in\Psi_k$ is closest to the typical user and therefore, all interfering satellites of type $k$ are at distances greater than $ d $. The set of interfering satellites of the type $k$ is denoted by 
\begin{equation}
		\bar{\Psi}_{k}^{u}\setminus X_k^\star= {\Psi}_k(C(r_k,\sqrt{r_k^2-r_e^2})\setminus C(r_k,z)).
\end{equation}
Using above notation, $\cL_{I_{\text{cl}, k}^{X_U}}(s)$ is given by 
 \begin{align}
	&\cL_{I_{\text{cl}, k}^{X_U}}(s)\nnb\\
	&=\prod_{Z_i\in{\Xi}_{k}}\bE\left[\prod_{X_j\in \bar{\psi}_{{i}}\setminus X_U} e^{-sp_{{k}} H\|X_j-u\|^{-\alpha}}|\Xi_k,Z_\star,d\right]\nnb\\
	&=\prod_{Z_i\in{\Xi}_{k}+\delta_{Z_\star}}^{|\phi_i-\pi/2|<\xi_k}e^{-\frac{\mu_{k}}{\pi}\int_{\omega_{k,\phi_i,1}}^{\omega_{k,\phi_i,2}}\left(1-\cL_H(sp_k/(\widetilde{f}_{k,\phi_i}^\alpha(\omega)))\right)\diff \omega}\nnb\\
	&\hspace{4mm}\prod_{Z_i\in{\Xi}_{k}}^{\xi_k<|\phi_i-\pi/2|<\bar{\varphi}_k}e^{-\frac{\mu_{k}}{\pi}\int_0^{\omega_{k,\phi_i,2}}\left(1-\cL_H(sp_k/(\widetilde{f}_{k,\phi_i}^\alpha(\omega)))\right)\diff \omega}\label{eq:144},
\end{align}
where we use the change of variable to get the integration w.r.t. $\omega.$ In Eq. \eqref{eq:144}, we use 
\begin{align}
	\cos(\xi_k)&=(r_k^2+r_e^2-z^2)/(2r_kr_e),\\
	\omega_{k,\phi_i,1}&=\arcsin(\sqrt{1-\cos(\xi_k)^2\csc^2(\phi_i)}),\\
	\omega_{k,\phi_i,2}&= \arcsin(\sqrt{1-(r_e/r_k)^2\csc^2(\phi_i)}),\\
		\tilde{f}_{k,\phi_i}(\omega_j)&=\sqrt{r_k^2-2r_kr_e\cos(\omega_j)\sin(\phi_i)+r_e^2}.
\end{align}
where $z$ denotes the distance from the typical user to the association satellite. 
\par Using Eq. \eqref{eq:SINR} and Fubini's theorem, the coverage probability of the type $k$ typical user is 
\begin{align}
	\bP(\SINR_{\textnormal{cl},k}\geq\tau)
	&=\int_0^{}e^{-\sigma^2v}\bE_{\Xi_k}\left[\bE_{Z_\star}\left[\cL_{I_{cl}}(v) h_{c}(z)\right]\right]\diff z,\label{22}
\end{align}
where $ v = \tau z^\alpha/(p_kg_k) $. Above, $ h_c(z) $ is the PDF of the distance from the typical user to its nearest type $k$ LEO satellite  evaluated at $z$, conditionally on $ \Xi_k $ and $Z_\star.$ Conditionally on $\Xi_k$  and on the event that $X_U$ is on the orbit given by the point  $Z_{\star}=(r_k,\theta_\star,\varphi_{\star})$,  the PDF is given by 
\begin{align}
	h_{c}(z) =&\partial_z \bP(\|X_k^\star-U\|\leq z |\Xi_k,Z_\star )\nnb\\
	=&\partial_{z}\left(1-\bP(X_j>z,\forall X_j\in Z_\star)\right)\nnb\\
	& \times \bP_{\Xi_k}^{!Z_\star}(X_j>z,\forall X_j\in Z_i,\forall Z_i\in\Xi_k ) \nnb\\
	=&\frac{\mu_{k} z|\csc(\phi_\star)|e^{-\frac{\mu_k}{\pi}\arcsin(\sqrt{1-\cos^2(\xi_k)\csc^2(\phi_\star)})}}{\pi r_kr_e\sqrt{1-\cos^2(\xi_k)\csc^2(\phi_\star)}}\nnb\\
	&\prod_{Z_i\in{\Xi}_{k}}^{|\phi_i-\pi/2|<\xi_k} e^{-\frac{\mu_{k}}{\pi}\arcsin(\sqrt{1-\cos^2(\xi_k)\csc^2(\phi_i)})}. \label{23}
\end{align}
To get the final result, we first combine Eqs. \eqref{eq:18}, \eqref{eq:144}, \eqref{22}, and \eqref{23}. Then we change the integration order to integrate w.r.t. $Z_\star $ first. Then we use the Campbell mean value theorem to integrate w.r.t. $Z_\star=(\theta_\star,\varphi_{\star})$ and then employ the probability generating functional of $\Xi.$ As a result, we obtain the final result given by Eq. \eqref{eq:theorem:1} where 
\begin{align}
				\cos(\xi_k) &=  (r_k^2+r_e^2-z^2)/(2r_kr_e)\label{p3},\\
		\omega_{k,\varphi,1}&=\arcsin(\sqrt{1-\cos(\xi_k)^2\sec^2(\varphi)})\label{p1},\\
		\omega_{k,\varphi,2}&= \arcsin(\sqrt{1-(r_e/r_k)^2\sec^2(\varphi)})\label{p2},\\
			\overbar{f}_{l,\varphi}(\omega)&=\sqrt{r_l^2-2r_lr_e\cos(\omega)\cos(\varphi)+r_e^2}\label{p4}.
\end{align}
This completes the proof. 
\end{IEEEproof}

\begin{figure}
	\centering
	\includegraphics[width=1.0\linewidth]{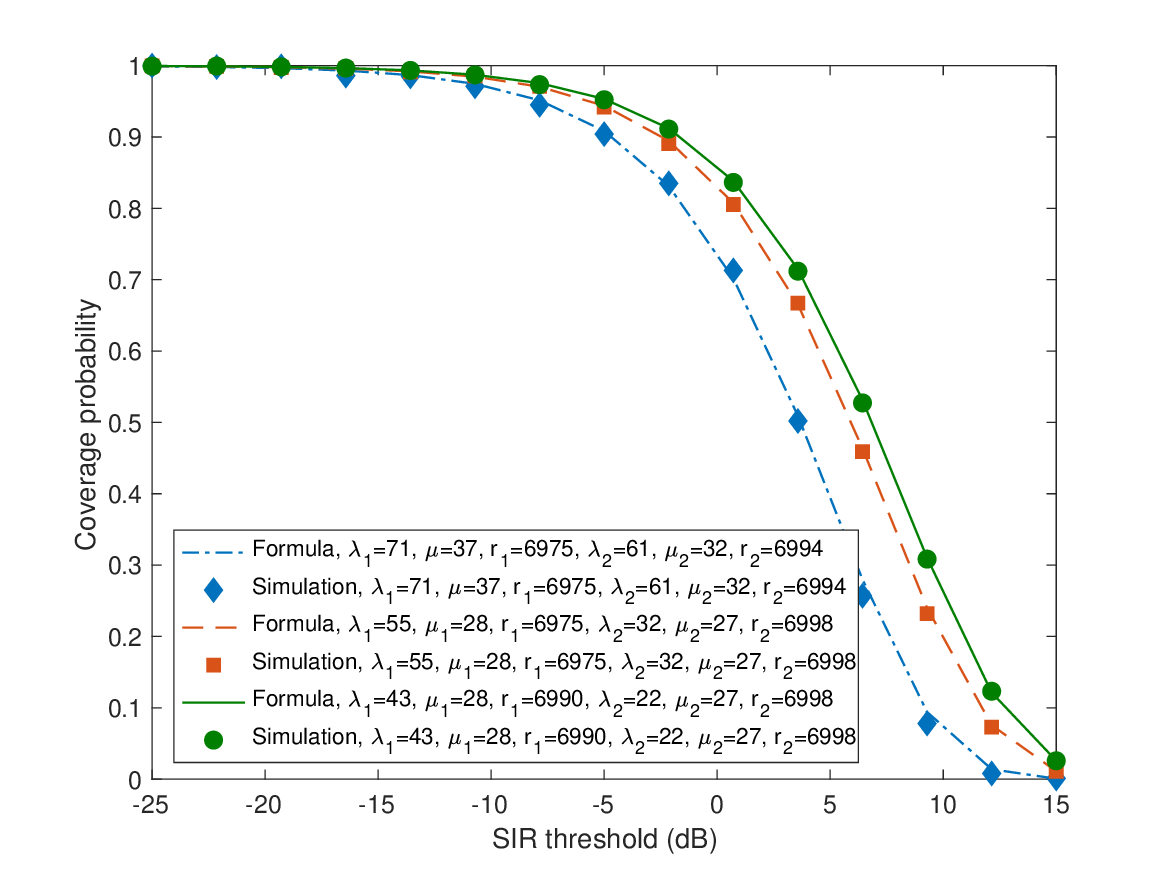}
	\caption{The coverage probability of the type $k=1$ typical user obtained by analysis and simulation, respectively.}
	\label{fig:closedexpcombined}
\end{figure}

\begin{figure}
	\centering
	\includegraphics[width=1.0\linewidth]{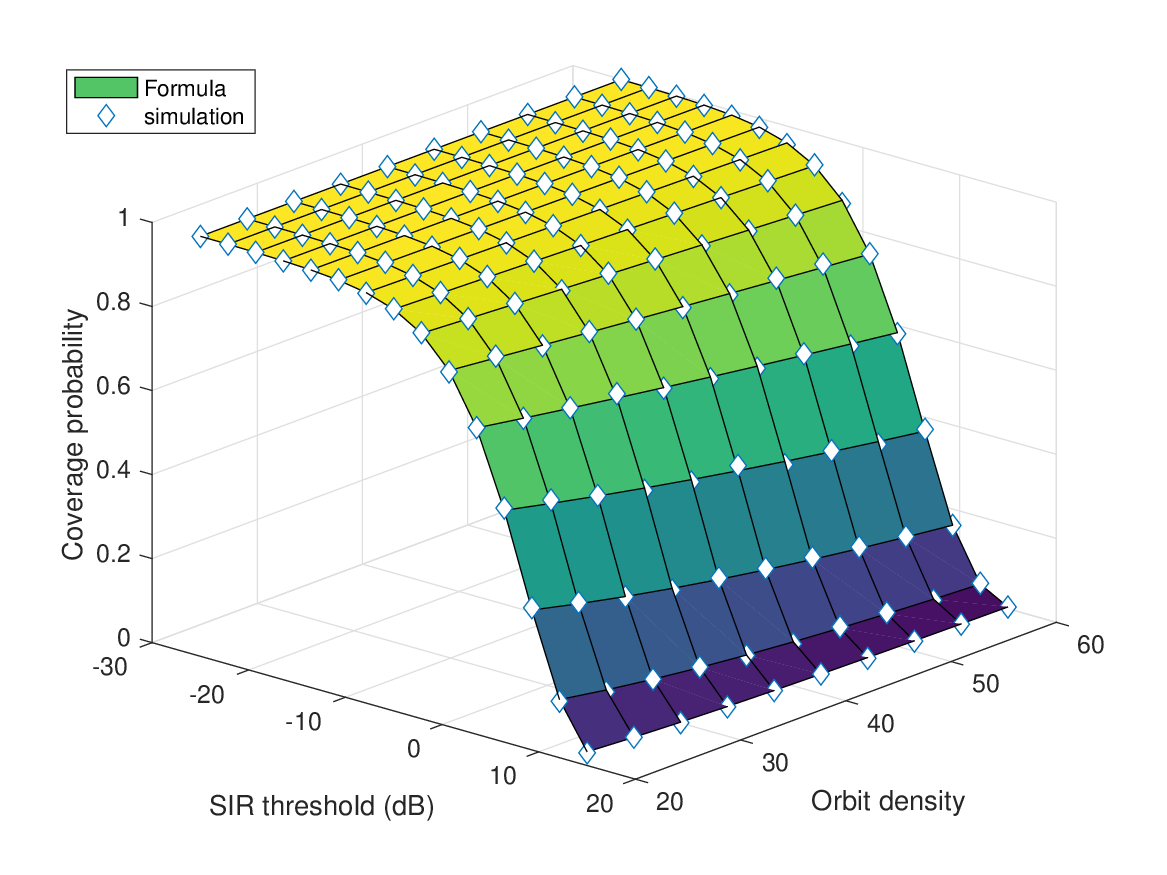} 
	\caption{The closed access coverage probability for $K=2, \lambda_1=40,\mu_1=\mu_2=30, r_1=r_2 = 6950$ km, and $g_1=20$ dB.}
	\label{fig:closedaccesschangevariable}
\end{figure}

\begin{figure}
	\centering
	\includegraphics[width=1.0\linewidth]{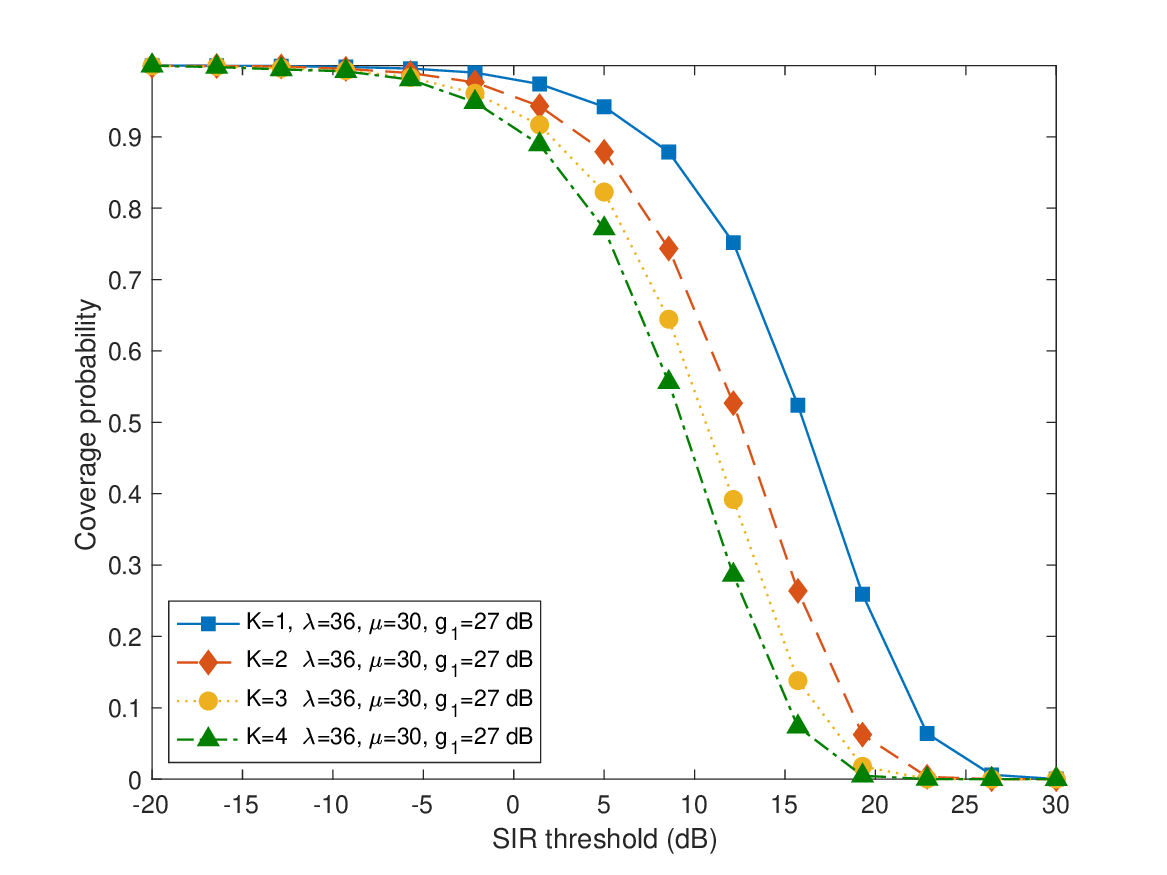}
	\caption{The closed access coverage probability for various $ K= 1,2,3,4$.}
	\label{fig:theorem1_3_2}
\end{figure}

\begin{figure}
	\centering
	\includegraphics[width=1.0\linewidth]{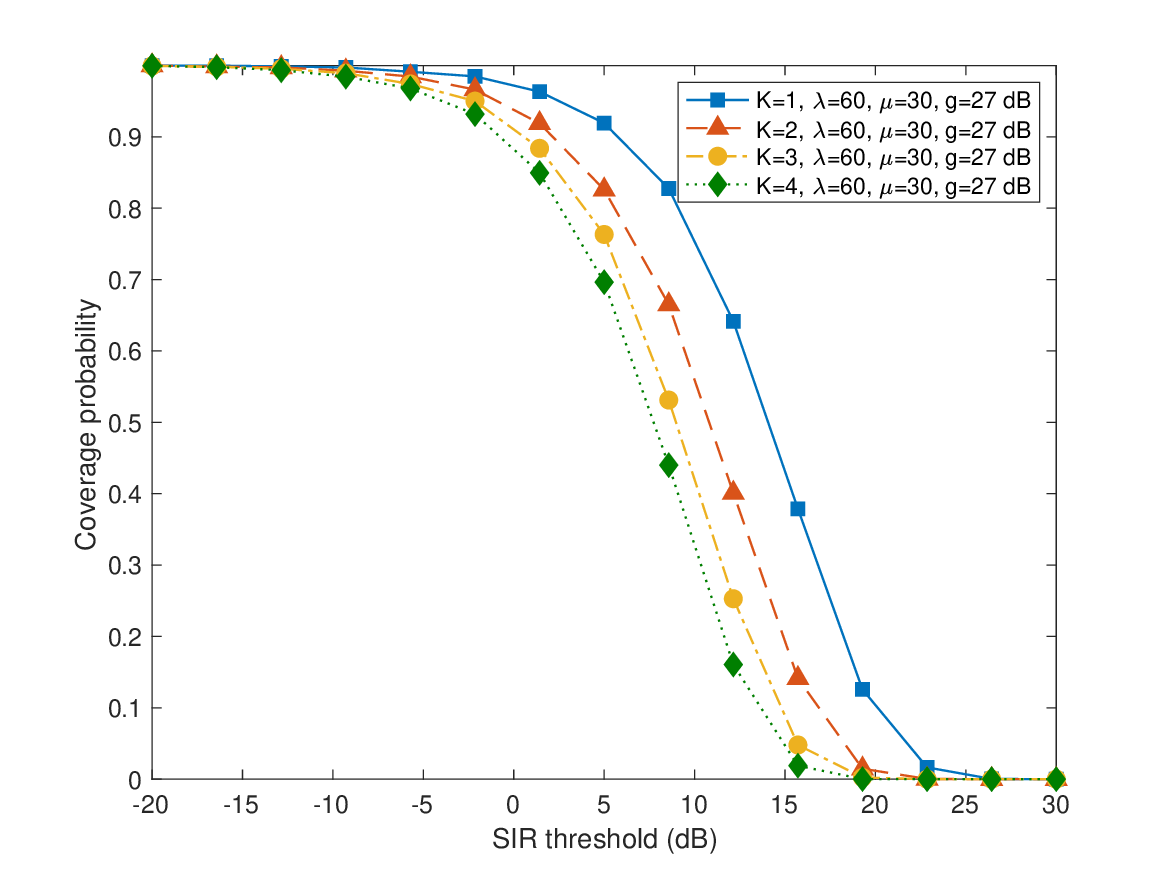}
	\caption{The closed access coverage probability for various $ K= 1,2,3,4$.}
	\label{fig:theorem1_3_3}
\end{figure}

Fig. \ref{fig:closedexpcombined} shows the coverage probability of the type $k$ typical user with $K=2$ types of constellations.  We test for various values of $\lambda_1, \lambda_2,\mu_1,\mu_2,r_1$, and $r_2$, to investigate the behavior of the coverage probability of the type $k$ typical user in the closed access scenario. The simulation results confirms the derived formula.

Fig \ref{fig:closedaccesschangevariable} displays the coverage probability of the type $k$ typical user as we increase or decrease $\lambda_2 $ the number of orbital planes of the interfering constellation. The figure illustrates that as the number of orbital planes increases, so does the total interference seen by the typical user. In Fig. \ref{fig:closedaccesschangevariable}, the radii of orbits are fixed to display the variation of the coverage probability with $\lambda_2$. For various orbit densities, the accuracy of the derived formula is validated by Monte Carlo simulation results. 

Fig. \ref{fig:theorem1_3_2} and \ref{fig:theorem1_3_3} depict the coverage probability of type $k$ user as total number of interfering constellations increases from $ K=2 $ to $ K=4$. This experiment specifically accounts for future LEO satellite deployment scenario where the demands and use cases for LEO satellite downlink communication is high and the number of LEO network operators is also large. We use $g_k=27$ dB, the mean number of orbits $ \lambda=36, $ and the mean number of satellites per orbit $ \mu=30 $ in Fig. \ref{fig:theorem1_3_2}. In $ K=1 $, there are no other interfering constellation and thus only inter-satellite interference of the same constellation exists. In $ K=4 $ there are three other interfering constellations. We observe that as $k$ increases, the coverage probability of the typical user of type $1$ dramatically decreases since in the closed access heterogeneous satellite networks, the association LEO satellite is not necessarily the closest one from the typical user.  
\begin{example}
From our experiments, we observe that as more satellites appear, the coverage probability rapidly decreases. Independently deployed constellations can not cooperate, and therefore the interference will continue to grow as well. In the following section, we study an open access scenario where users are enabled to communicate with any type of satellite. In this scenario, users are associated with the closest LEO satellite regardless of its type, and therefore some of the local interference manifest in the closed access system will vanish. We will derive the coverage probability of the typical user in the open access system and then compare it to the coverage probability of the typical user in the closed access system to showcase the benefit of the open access LEO satellite constellation system.  
\end{example}
\begin{corollary}\label{C:2}
	Under the closed access scenario, the ergodic capacity of the typical downlink communication of type $k$ is
	\begin{equation}
		\int_{0}^\infty \bP(\SINR_{\textnormal{cl},k}>2^r-1 ) \diff r 
	\end{equation}
where $\bP(\SINR_{\textnormal{cl},k}>\cdot)$ is given by Eq. \eqref{eq:theorem:1} of Theorem \ref{Theorem:1}.
\end{corollary}
\begin{IEEEproof}
Let $\log(1+\SINR_{\textnormal{cl},k})$ represent the ergodic capacity of the typical downlink communication of type $k.$ Due to the rotation invariance of $\Psi$ and the independence of the user point process and the satellite point process, the aforementioned SINR random variable corresponds to the SINR observed by the typical user.

 Therefore, we integrate its probability distribution function to arrive at the following expression \begin{align}
			\bE[\log_2(1+\SINR_{\textnormal{cl},k})]&= \int_{0}^{\infty} \bP(\log_2(1+\SINR_{\textnormal{cl},k})>r) \diff r \nnb\\&=  \int_{0}^{\infty} \bP(\SINR_{\textnormal{cl},k}>2^r-1) \diff r,\label{zz}
		\end{align}
		where the integrand of Eq. \eqref{zz} is the coverage probability given by Eq. \eqref{eq:theorem:1}, evaluated at $2^r-1$ .  
\end{IEEEproof}
\section{Open Access}\label{S:4}
In this section, we analyze the proposed heterogeneous LEO satellite network where users can communicate with their closest LEO satellites, regardless of the types of users or LEO satellites. Thanks to this open access, we expect that the coverage probability of the typical user will increase. 
\subsection{Open Access No-Satellite Probability}
The following example highlights the benefits of the open access, specifically in the context of no-satellite probability. Below, we examine the open access no-satellite probability in the proposed heterogeneous satellite network. 
\begin{proposition}\label{prop:2}
	The open access no-satellite probability in the proposed heterogeneous satellite network is 
	\begin{align}
				\prod_{k=1}^K&\exp\!\left(-{\lambda_{k}}\!\int_0^{\arccos({r_e}/{r_k})}\right.\nnb\\
				&\hspace{10mm}\left.\left(1-e^{-\frac{\mu_{k}}{\pi} \arcsin(\!\sqrt{1-r_e^2\sec^2(\varphi)/r_{k}^2})}\right)\cos(\varphi)\!\diff \varphi\!\right).
	\end{align}
\end{proposition}
\begin{IEEEproof}
In the open access, users can communicate with any type of satellites. Thus, no-satellite occurs if and only if the typical user is unable to see any satellite of any type. Since all $K$ LEO satellite constellations are spatially independent, we have 
\begin{equation}
\bP(\textnormal{no-satellite}) = \prod_{k=1}^{K}\bP(\textnormal{no-satellite}_k),
\end{equation}
where $\bP(\textnormal{no-satellite}_k)$ is the closed access no-satellite probability of type $k$ in Proposition  \ref{prop:1}.  
\end{IEEEproof}

Next, we evaluate the probability that users are associated with a certain type of LEO satellites. It relates to the amount of network users that each LEO satellite constellation can serve in the open access scenario. Note we exploit the association probability to obtain the SINR coverage probability of the typical user. 

\begin{figure*}
	\begin{align}
		&\int_{r_k-r_e}^{\sqrt{r_k^2-r_e^2}} {\lambda_{k}  g_k(v)}
		\left(\prod_{m=1}^{K}\left\{ 1\ind_{v< r_m-r_e}+F_{D_m}(v)\ind_{r_{m}-r_{e}\leq v< \sqrt{r_{m}^2-r_{e}^2}}+ \bP(\textnormal{no-satellite}_{m})\ind_{v\geq \sqrt{r_m^2-r_e^2}}\right\}\right)\diff v\label{eq:Theorem:1-1}.
	\end{align}
	\rule{\linewidth}{0.2mm}
\end{figure*}

\begin{theorem}\label{Theorem:2}
In a case that $r_i\neq r_j, \forall i\neq j$, the association probability $\bP(\cA_k)$ is given by Eq. \eqref{eq:Theorem:1-1}. 

If $r_1=\cdots=r_K,$ the association probability  $\bP(\cA_k)$  is 
	\begin{equation}
		\int_{r_k-r_e}^{\sqrt{r_k^2-r_e^2}} {\lambda_{k}g_k(v)}\left(\prod_{m=1}^{K} F_{D_m}(v)\right)\diff v
	\end{equation}
	where $g_k(v)$ is in Eq. \eqref{gkv}. 
\end{theorem}
\begin{IEEEproof}
The typical user is associated with a type $ k $ LEO satellite if the nearest distance from $ U $ to the LEO satellites of type $ k $ is less than the nearest distance from $ U $ to the LEO satellites of types $ [K]\setminus k. $ The association probability $ \bP(\cA_{k}) $ is 
	\begin{align}
		\bP(\cA_{k}) & = \bP(D_m>D_k, \forall m\in[K]\setminus k), \nnb
	\end{align}
where $ D_k $ denotes the distance from $ U $ to its closest type $ k $ LEO satellite. Conditionally on $ D_k=v, $ we get 
\begin{align}
	\bP(\cA_{k}) & = \bE_{D_k}\left[\bP(D_m> v, \forall m\in[K]\setminus k|D_k =v)\right].\nnb
\end{align}
Since  $ \Psi_1, \Psi_2,...,\Psi_K $ are independent, we get 
\begin{align}
	\bP(\cA_{k}) & = \bE_{D_k}\left[\prod_{{m=1},{m\neq k}}^{K}\bP(D_m> v)\right].\nnb
\end{align}
We use Lemma \ref{Lemma:5} to arrive at the following CCDF expression
\begin{align}
	&F_{D_m}(v) \nnb\\
	&=\bP(D_m>v) \nnb\\&= \begin{cases}
		1 , \text{ for }v<r_m-r_e \\ 
		\exp\left(-{\lambda_m}\int_0^{\xi_m}\!\!\cos(\varphi)\right.\\
	\hspace{15mm}	\left.\left(1-e^{\!-\frac{\mu_{m}}{\pi}\sin^{-1}\left(\sqrt{1-\!\frac{\cos^2(\xi_m)}{\cos^2(\varphi)}}\right)}\right)\diff \varphi \right),\\
				\bP(\textnormal{no-satellite}_m)  \text{ for } v>\sqrt{r_m^2-r_e^2},\nnb
	\end{cases}
\end{align} 
where $\xi_m = \arccos((r_m^2+r_e^2-v^2)/(2r_mr_e))$ and 
\begin{align}
&\bP(\textnormal{no-satellite}_m)\nnb\\
&=		e^{-{\lambda_{m}}\int_0^{\overbar{\varphi}_m}\cos(\varphi)\left(1-e^{-\frac{\mu_{m}}{\pi} \arcsin\left(\sqrt{1-r_e^2\sec^2(\varphi)/r_{m}^2}\right)}\right)\diff \varphi}.\nnb
\end{align}

Meanwhile, the PDF of $ D_k $ is given by 
\begin{align}
	f_{D_k}(v)
	&=\frac{\partial }{\partial v}\left\{1-\exp\left(-{\lambda_k}\int_0^{\xi_k}\cos(\varphi)\right.\right. \nnb\\ 
	&\hspace{15mm}\left.\left.\left(1-e^{-\frac{\mu_{k}}{\pi}\arcsin\left(\sqrt{1-\frac{\cos^2(\xi_k)}{\cos^2(\varphi)}}\right)}\right)\diff \varphi \right)\right\}\nnb\\
	&={\lambda_{k}}F_{D_k}(v)g_k(v),
\end{align}
where 
\begin{align}
	g_k(v)=&\frac{\mu_{k} v}{\pi r_e r_k}\int_0^{\xi_{k}}\frac{e^{-\frac{\mu}{\pi}\arcsin\left(\sqrt{1-{\cos^2(\xi_k)}{\sec^2(\varphi)} }\right)}}{\sqrt{1-\cos^2(\xi_k)\sec^2(\varphi) }}\diff \varphi,\label{gkv}
\end{align} 
and $\cos(\xi_k) = (r_k^2+r_e^2-v^2)/(2r_kr_e). $
Finally, we use the CCDF and PDF expressions above to get the final result.
\end{IEEEproof}
\begin{figure}
	\centering
	\includegraphics[width=1.0\linewidth]{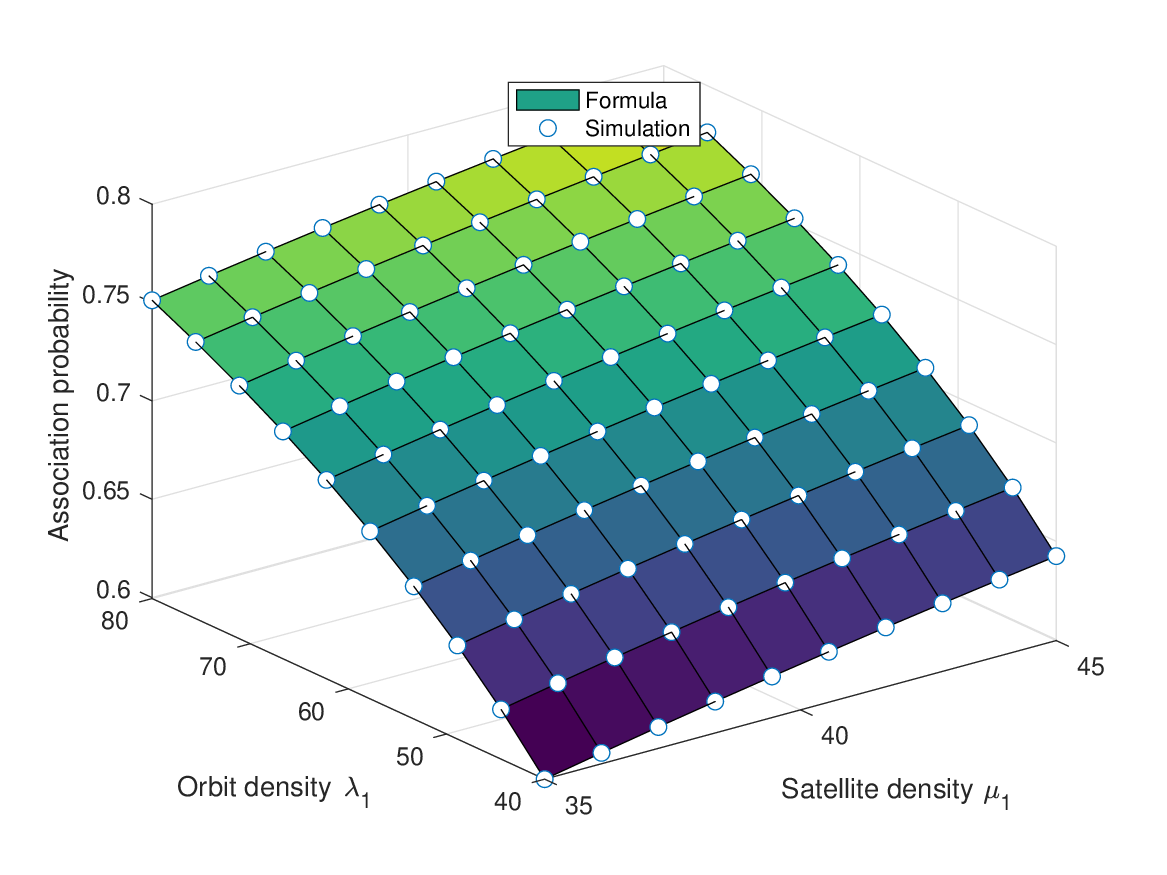}
	\caption{The association probability $\bP(\cA_1)$ for various $\lambda_1$ and $\mu_1 $. We use $\lambda_2=\mu_2=30,$ and  $r_1=r_2=7000 $ km.}
	\label{fig:associationvariablelambdamu}
\end{figure}

\begin{figure}
	\centering
	\includegraphics[width=1.0\linewidth]{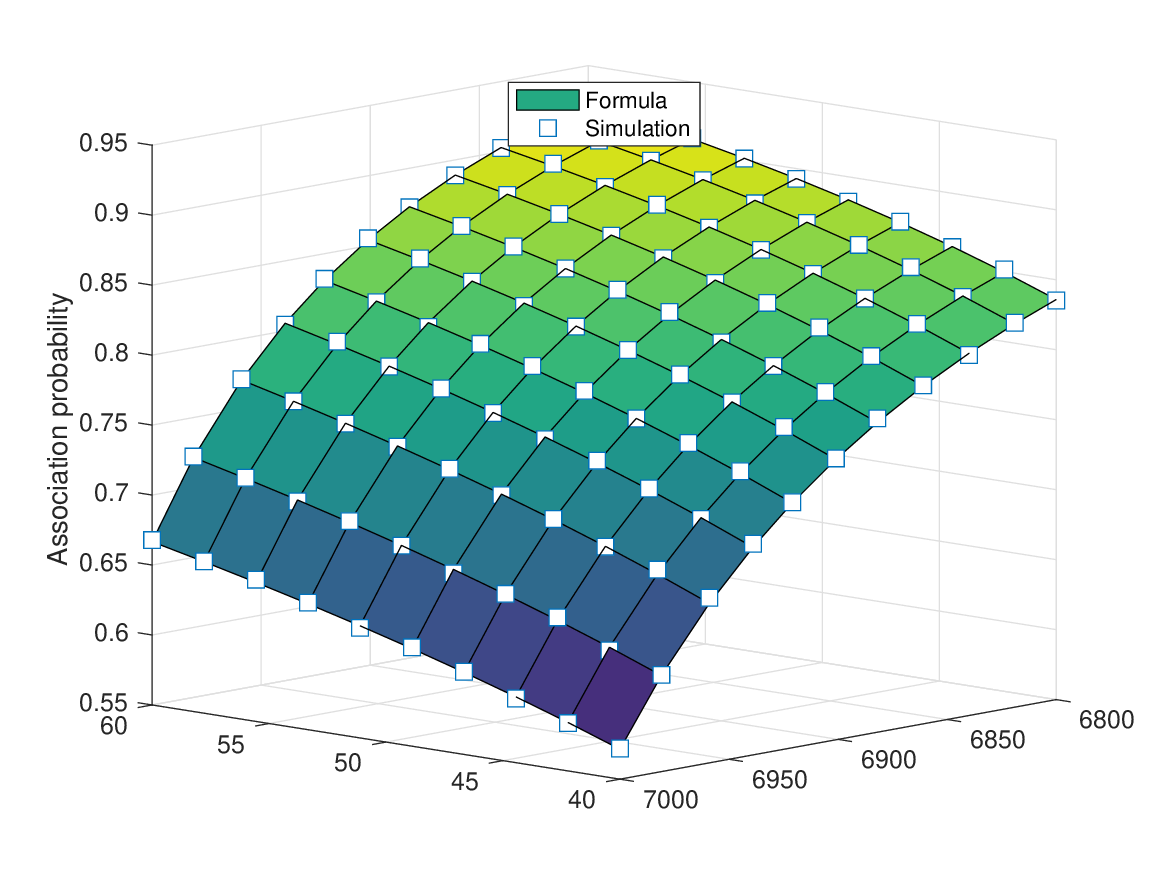}
	\caption{The association probability $\bP(\cA_1)$ for various $\lambda_1$ and $r_1$. We use $\mu_1=\lambda_2=\mu_2 = 30, $ and  $ r_2 = 7000$ km.}
	\label{fig:associationvariableradius}
\end{figure}

Figs. \ref{fig:associationvariablelambdamu}  and \ref{fig:associationvariableradius} show $\bP(\cA_1)$ the association probability that the typical user is associated with the type $1$ satellite. The simulation results validate the accuracy of the derived formula in Theorem \ref{Theorem:2}.  We use $K=2 $ for both figures. In Fig. \ref{fig:associationvariablelambdamu}, we vary $\lambda_1$ and $\mu_1$ to see the probability that the typical user is associated with the satellite of type $1.$  Here, we use $\lambda_2=30$ and $\mu_2 = 30$ for $\Psi_2.$ We observe that as $\lambda_1 $ or $\mu_1$ increases, the typical user is more likely to be associated with the satellite of type $1.$ Fig. \ref{fig:associationvariableradius} illustrates the behavior of the association probability as $r_1$ and $\lambda_1 $ vary.  It is important to emphasize that the computational expense associated with conducting simulations is significantly higher than that of utilizing the derived formula. To get smoother results, one needs more than $N=10^7 $ samples for simulations. 

The data presented in the figure suggests that, as the altitude of type $1$ satellites decreases, the typical user tends to be associated with such satellites. This association holds for all values of $\lambda_{1}$. It is evident from this information that a lower-altitude satellite constellation accommodates a larger number of network users, compared to a higher-altitude satellite constellation. This conclusion is based solely on the distances-based association principle that we have assumed in Section \ref{S:2-a}. Network designers can derive valuable insights from this observation regarding the distribution and deployment of satellites across orbital planes to effectively control and serve the network users in heterogeneous satellite networks.

	Instead of the nearest distance association, network operators may employ the maximum average receive signal power association incorporating not only (i) the distances to LEO satellites and corresponding large-scale attenuation but also (ii) array gains from transmit and receive antennas, for the best coverage experience. Leveraging Theorem \ref{Theorem:2}, we can also derive the association probability of the typical user in such a scenario. Now, the association probability $ \bP(\overbar{\cA_{k}}) $ is now  
	\begin{align}
		\bP(\overbar{\cA_{k}}) & = \bP(p_{k}g_kD_k^{-\alpha}> p_{m}g_mD_m^{-\alpha}, \forall m\in[K]\setminus k) \nnb\\
		&=\bP\left(D_{m}\geq \sqrt[\alpha]{\eta_{m,k}} D_k, \forall m\in[K]\setminus k\right)\nnb\\
		&=\bE_{D_{k}}\left[\bP(\forall m\in[K]\setminus k, D_m> v\sqrt[\alpha]{\eta_{m,k}}|D_k=v)\right],\nnb
	\end{align}
	where we let  $ \eta_{m,k} = (p_mg_m)/(p_kg_k). $  
	
	Using the proof of Theorem \ref{Theorem:1}, we obtain the following 
	\begin{equation}
		\int_{r_k-r_e}^{\sqrt{r_k^2-r_e^2}} {\lambda_{k}g_k(v) F_{D_k}(v)}\left(\prod_{m\in[K]\setminus k} F_{D_m}(v\sqrt[\alpha]{\eta_{m,k}})\right) \diff v\nnb.
	\end{equation}

\subsection{Open Access Coverage Probability}
In below, we evaluate the coverage probability of the typical user. We use the fact that given the user association, the LEO satellite closest to the typical user is the association LEO satellite transmitter. Therefore, all the other LEO satellites of any type are located at distances further than the association LEO satellite. It is expected that when every network parameter is the same, the open access coverage probability is greater than the closed access coverage probability. 
 
\begin{theorem}\label{Theorem:3}
When $m=1$, the coverage probability of the typical user in the proposed open access $ K $ multiple LEO satellite constellations is given by Eq. \eqref{eq:theorem:3} whereas $\bP(\cA_k)$ is given by Eq. \eqref{eq:Theorem:1-1}. 
\end{theorem}
\begin{figure*}
	\begin{align}
		&\sum_{k=1}^K \int_{r_k-r_e}^{\sqrt{r_k^2-r_e^2}}\cA_ke^{-\frac{\sigma^2 \tau z^\alpha}{p_kg_k}}\frac{\lambda_k \mu_k z }{\pi r_kr_e}\nnb\\
				&\hspace{21mm}\left(\prod_{l=1}^{K}\exp\left(-{\lambda_{l}}{}\int_{0}^{{\xi}_l}\left(1- e^{-\frac{\mu_l}{\pi}\arcsin(\sqrt{1-\cos(\xi_l)^2\sec(\varphi)^2})-\frac{\mu_{l}}{\pi}\int_{\omega_{l,\varphi,1}}^{\omega_{l,\varphi,2}}\left(1-\cL_H(\tau z^\alpha  g_k^{-1}\overbar{f}_{l,\varphi}^{-\alpha}(\omega))\right)\diff \omega}\right)\cos(\varphi)\diff \varphi\right)\right)\nnb\\
				&\hspace{21mm}\left(\prod_{l=1}^{K}\exp\left(-{\lambda_{l}}\int_{\xi_{l}}^{\bar{\varphi}_l}\left(1- e^{-\frac{\mu_{l}}{\pi}\int_{\omega_{l,\varphi,1}}^{\omega_{l,\varphi,2}}\left(1-\cL_H(\tau z^\alpha  g_k^{-1}\overbar{f}_{l,\varphi}^{-\alpha}(\omega))\right)\diff \omega}\right)\cos(\varphi)\diff \varphi\right)\right)\nnb\\  &\hspace{21mm}\left(\int_{0}^{w}\frac{e^{-\frac{\mu_k}{\pi}\arcsin(\sqrt{1-\cos(\xi_k)^2\sec(v)^2})-\frac{\mu_{k}}{\pi}\int_{\omega_{k,v,1}}^{\omega_{k,v,2}}\left(1-\cL_H(\tau z^\alpha  g_k^{-1}\overbar{f}_{k,v}^{-\alpha}(\omega))\right)\diff \omega}}{\sqrt{1-\cos^2(\xi_k)\sec^2(v)}}\diff v\right)  \diff z \label{eq:theorem:3}.
	\end{align}
	\rule{\linewidth}{0.1mm}
\end{figure*}

\begin{IEEEproof}
	The SINR coverage probability is given by 
	\begin{align}
		\bP(\SINR_{\text{op}}\geq \tau ) = \sum_{k=1}^K \bP(\SINR_{\text{op}}\geq \tau | \cA_k)\bP(\cA_k).
	\end{align}
where $ \cA_k $ denotes an event that the association LEO satellite is of type $ k. $ The coverage probability of the typical user, conditionally on this event, is given by 
\begin{align}
	&\bP(\SINR_{\text{op}}\geq \tau | \cA_k) \nnb\\
	& = \bP\left(\frac{p_{{k}} g_kH }{ (\sigma^2+I_{\text{op}}) \|X_k^\star-U\|^{\alpha}}\geq \tau \right)\nnb\\
	&=\bE\left[\bE\left[\bE\left[\bP\left(\left.H\geq\frac{\tau (\sigma^2+I_{\text{op}})z^{\alpha}}{p_kg_k}\right| \Xi,Z_\star , d\right) \right]\right]\right]\nnb\\
	&=\bE\left[\bE\left[\bE\left[ e^{-\frac{\tau\sigma^2z^\alpha}{p_kg_k}}\cL_{I_{\text{op}}}\left( \frac{\tau z^\alpha}{p_kg_k}\right)\right]\right]\right]\label{eq:SINR2}.
\end{align}
In above, $ X_k^\star $ is the LEO satellite of type $ k $ nearest from $ U $ and $ Z_\star $ corresponds to the orbit that contains $ X_k^\star $. We use $ d=\|X_k^\star-U\| $ and $ \cL_{I_{\text{op}}} (s) $ is the Laplace transform of the interference seen by the typical user at $ U. $ We have 
\begin{align}
	I_{\text{op}}&=\sum_{l=1}^K I_{\text{op},l}^{X_k^\star}, \\ 
	\cL_{I_{\text{op}}}(s)&=\prod_{l=1}^K\cL_{I_{\text{op}, l}^{X_k^\star}}(s),
\end{align}
where $ I_{\text{op},l}^{X_k^\star} $ is the interference generated by the type $ l $ visible satellites located further than the association satellite. Let $ \bar{\Psi}_{{l}}^u $ be the set of such satellites and we have 
\begin{equation}
		\bar{\Psi}_{{l}}^u \setminus X_k^\star \equiv \Psi_{{l}}(C(r_l,\sqrt{r_l^2-r_e^2})\setminus C(r_l,z)). 
\end{equation}
Then, the Laplace transform of the interference from type $ l $ satellites are given by 
\begin{align}
	I_{\text{op},l}&=\sum_{X_j\in \bar{\Psi}_{{l}}^u\setminus X_k^\star
		} p_l H\|X_j- U \|^{-\alpha} .
\end{align}
For $ l\neq k, $ $ \Psi_l\independent \{\Psi_k,\Xi_k,Z_\star\} $. 

Conditionally on $\Xi,Z_\star,$ and $z,$ the Laplace transform of the interference from LEO satellites of type $l$ is given by 
\begin{align}
		&\cL_{I_{\text{op}, l}^{X_k^\star}}(s)\nnb\\
		&=\bE\left[e^{-s\sum_{X_j\in \bar{\Psi}_{{l}}^u \setminus X_k^\star} p_{{k}} H\|X_j-U\|^{-\alpha}}|\Xi,Z_\star,d\right]\nnb\\
		&=\prod_{Z_i\in{\Xi}_{l}}^{|\phi_i-\pi/2|<\xi_l}e^{-\frac{\mu_{l}}{\pi}\int_{\omega_{l,\phi_i,1}}^{\omega_{l,\phi_i,2}}\left(1-\cL_H(sp_l/\widetilde{f}_{l,\phi_i}^{\alpha}(\omega))\right)\diff \omega}\nnb\\
			&\hspace{4mm}\prod_{Z_i\in{\Xi}_{l}}^{\xi_l<|\phi_i-\pi/2|<\bar{\varphi}_l}e^{-\frac{\mu_{l}}{\pi}\int_0^{\omega_{k,\phi_i,2}}\left(1-\cL_H(sp_l/(\widetilde{f}_{l,\phi_i}^{\alpha}(\omega)))\right)\diff \omega},
			\label{eq:37}
\end{align}
where $\cos(\overbar{\varphi}_l) = r_e/r_l$ and we have 
\begin{align}
	&\cos(\xi_l) = (r_l^2+r_e^2-z^2)/(2 r_lr_e)\label{35},\\
	&\omega_{l,\phi_i,1}=\arcsin(\sqrt{1-\cos(\xi_l)^2\csc^2(\phi_i)}),\\
	&\omega_{l,\phi_i,2}=\arcsin(\sqrt{1-(r_e/r_l)^2\csc^2(\phi_i)}),\\
	&\widetilde{f}_{l,\phi_i}(\omega_j)=\sqrt{r_l^2-2r_lr_e\cos(\omega_j)\sin(\phi_i)+r_e^2}.\label{38}
\end{align}
Conditionally on $\Xi,Z_\star,$ and $z,$ the Laplace transform of the interference from LEO satellites of type $k$ is given by
 \begin{align}
	&\cL_{I_{\text{op}, k}^{X_k^\star}}(s)\nnb\\
	&=\prod_{Z_i\in{\Xi}_{k}+\delta_{Z_\star}}^{|\phi_i-\pi/2|<\xi_k}e^{-\frac{\mu_{k}}{\pi}\int_{\omega_{k,\phi_i,1}}^{\omega_{k,\phi_i,2}}\left(1-\cL_H(sp_k/\widetilde{f}_{k,\phi_i}^\alpha(\omega))\right)\diff \omega}\nnb\\
	&\hspace{4mm}\prod_{Z_i\in{\Xi}_{k}+\delta_{Z_\star}}^{\xi_k<|\phi_i-\pi/2|<\bar{\varphi}_k}e^{-\frac{\mu_{k}}{\pi}\int_{0}^{\omega_{k,\phi_i,2}}\left(1-\cL_H(sp_k/\widetilde{f}_{k,\phi_i}^\alpha(\omega))\right)\diff \omega},\label{eq:1442}
\end{align}
whose variables are given by Eq. \eqref{35} -- \eqref{38} with $ l $ replaced by $ k, $ respectively.

Using Eq. \eqref{eq:SINR2} and Fubini's theorem, the coverage probability of the typical user, conditional on $\cA_k$ is given by 
\begin{align}
	\bP(\SINR_{\text{op}}\geq\tau|\cA_k)
	&=\int_0^{}e^{-\sigma^2v}\bE_{\Xi,Z_\star}\left[\cL_{I_{op}}(v) g_{o}(z)\right]\diff z,\label{24}
\end{align}
where $ v = \tau z^\alpha/(p_kg_k) $ and $\bE_{\Xi,Z_\star} (\cdot) $ is the expectation with respect to $\Xi$ and $Z_\star.$

Above, $ g_c(z) $ is the PDF of the distance from $ U $ to its nearest type $ k $ LEO satellite, conditionally on $ \Xi$ and $ Z_\star. $  We arrive at 
\begin{align}
	g_{o}(z) =&\partial_z \bP(\|X_k^\star-U\| \leq z |\Xi,Z_\star )\nnb\\
	=&\partial_{z}\left(1-\bP(X_j>z,\forall X_j\in Z_\star,  Z_k\in\Xi_k )\right)\nnb\\
	& \times \bP_{\Xi_k}^{!Z_\star}(X_j>z,\forall X_j\in Z_i,\forall Z_i\in\Xi_k ) \nnb\\
	& \times \prod_{l\neq k}\bP_{\Xi_l}(X_j>z,\forall X_j\in Z_i,\forall Z_i\in\Xi_l ) \nnb\\
	=&\frac{\mu_{k}z|\csc(\phi_\star)|e^{-\frac{\mu_k}{\pi}\arcsin(\sqrt{1-\cos^2(\xi_k)\csc^2(\phi_\star)})}}{\pi r_kr_e\sqrt{1-\cos^2(\xi_k)\csc^2(\phi_\star)}}\nnb\\
	& \times  \prod_{l=1}^K\left(\prod_{Z_i\in{\Xi}_{l}}^{|\phi_{i}-\pi/2|<\xi_l} e^{-\frac{\mu_l}{\pi}\arcsin(\sqrt{1-\cos^2(\xi_l)\csc^2(\phi_i)})}\right), \label{23'}
\end{align}
where $ \varphi_\star $ is the azimuth angle of the orbit that contains the $ X_k^\star $ and $ \cos(\xi_l)= (r_l^2+r_e^2-z^2)/(2 r_lr_e) $ and $ \xi_{l}< \bar{\varphi}_l $ for $ z<r_e. $ 

Finally, we combine Eq. \eqref{eq:37}, \eqref{eq:1442}, \eqref{24}, and \eqref{23'} and then use the change of variable $\varphi = \pi/2-\phi_i$ and to obtain the final result in Eq. \eqref{eq:theorem:3} 
where 
\begin{align}
		\cos(\xi_l) &=  (r_l^2+r_e^2-z^2)/(2r_lr_e),\\
	\omega_{l,\varphi,1}&=\arcsin(\sqrt{1-\cos(\xi_l)^2\sec^2(\varphi)}),\\
	\omega_{l,\varphi,2}&= \arcsin(\sqrt{1-(r_e/r_l)^2\sec^2(\varphi)}),\\
	\overbar{f}_{l,\varphi}(\omega)&=\sqrt{r_l^2-2r_lr_e\cos(\omega)\cos(\varphi)+r_e^2}.
\end{align}
This completes the proof. 
\end{IEEEproof}
	
	
		Fig. \ref{fig:siropenvsclosedtotalantenna20db}  illustrates the coverage probability of a typical user both for the closed access and open access scenarios. We compare them by varying $K$ from $1 $ to $4$ with $\lambda_k=36$ and $\mu_k=20$ for all types. The direct comparison of the coverage probability shows that the coverage probability of the typical user in the closed access system rapidly drops as the number of interfering satellites increases. For $K = 4$, the average of the open access system is almost twice that of the closed access system. Specifically, the 90th percentile of the closed access typical user is $-5$ dB, while the $90$-th percentile of the open access typical user is $-2.5$ dB, which is $2.5$ dB higher than that of the closed access scenario. It is important to note that this improvement of $2$ to $3$ dB is more pronounced to users at low percentiles.

	
In Fig. \ref{fig:siropenvsclosedtotalantenna23db}, we present the coverage probability for a typical user of type $1$ in the presence of local interfering satellites with $\lambda_k=60,\mu_k=25$ for all constellations. Here, $\lambda_k\mu_k$ represents the total average number of satellites for type $k$, resulting in each constellation having twice as many satellites than Fig. \ref{fig:siropenvsclosedtotalantenna20db}.

For the case of $K=4$, the total number of satellites in the system is $2880$ in Fig. \ref{fig:siropenvsclosedtotalantenna20db}, while the total number of satellites is $6000$ in Fig. \ref{fig:siropenvsclosedtotalantenna23db}. It is evident that as the number of $K$ increases, both open access and close access coverage probabilities decrease. Additionally, the difference between open access coverage probability and closed access probability grows with increasing $K$ simply because the number of total interference satellites increases as well. This suggests that, with the proliferation of different LEO satellite operators, adopting open access technology could serve as a viable solution to mitigate interference from diverse constellations.
		
	\begin{corollary}
The ergodic capacity (or achievable rate) of downlink communications of open access user in the proposed heterogeneous satellite network is given 
		\begin{equation}
			\int_{0}^\infty \bP(\SINR_{\textnormal{op}}>2^r-1)\diff r . 
		\end{equation}The proof is immediate from Corollary \ref{C:2}.
	\end{corollary}

	\begin{figure}
		\centering
		\includegraphics[width=0.9\linewidth]{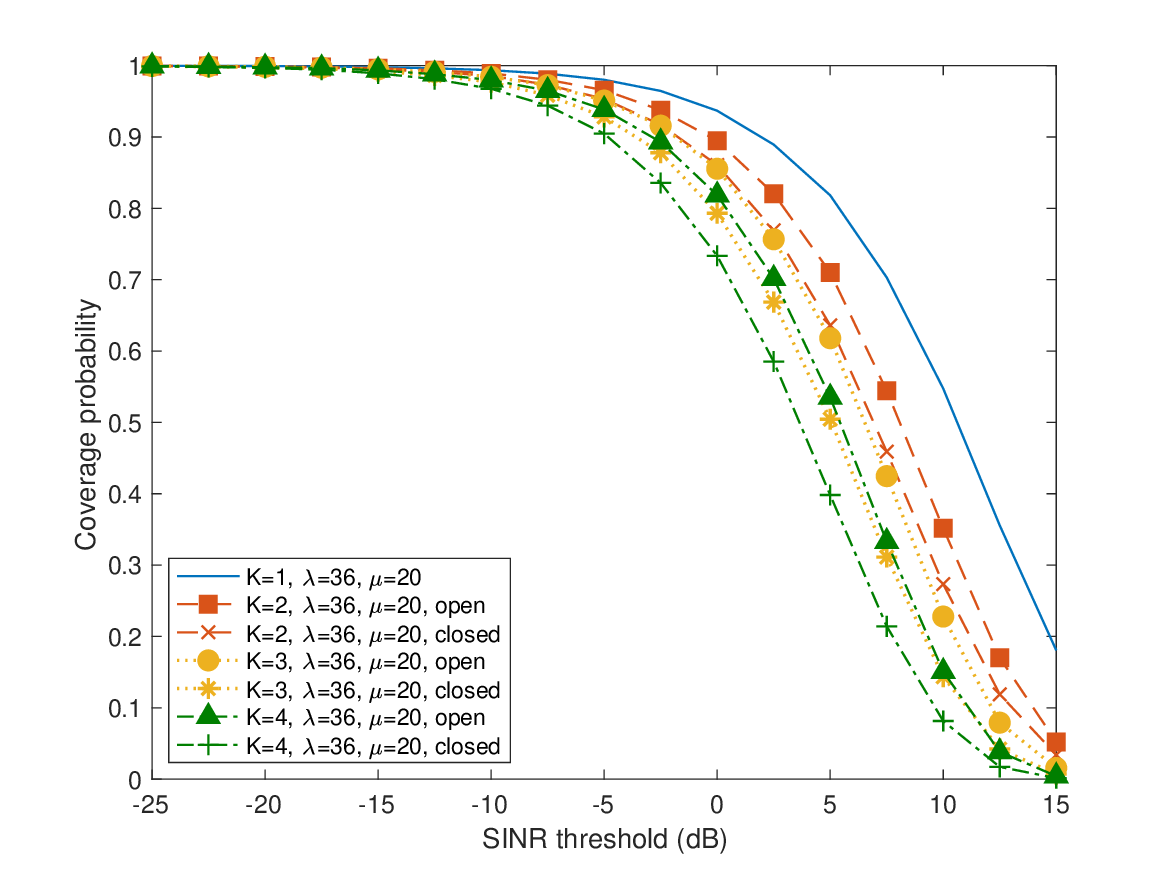}
		\caption{Illustration of the open access coverage probability. For all types, the satellite altitude is  $550$ km. $g=20$ dB.}
		\label{fig:siropenvsclosedtotalantenna20db}
	\end{figure}

	\begin{figure}
		\centering
		\includegraphics[width=0.9\linewidth]{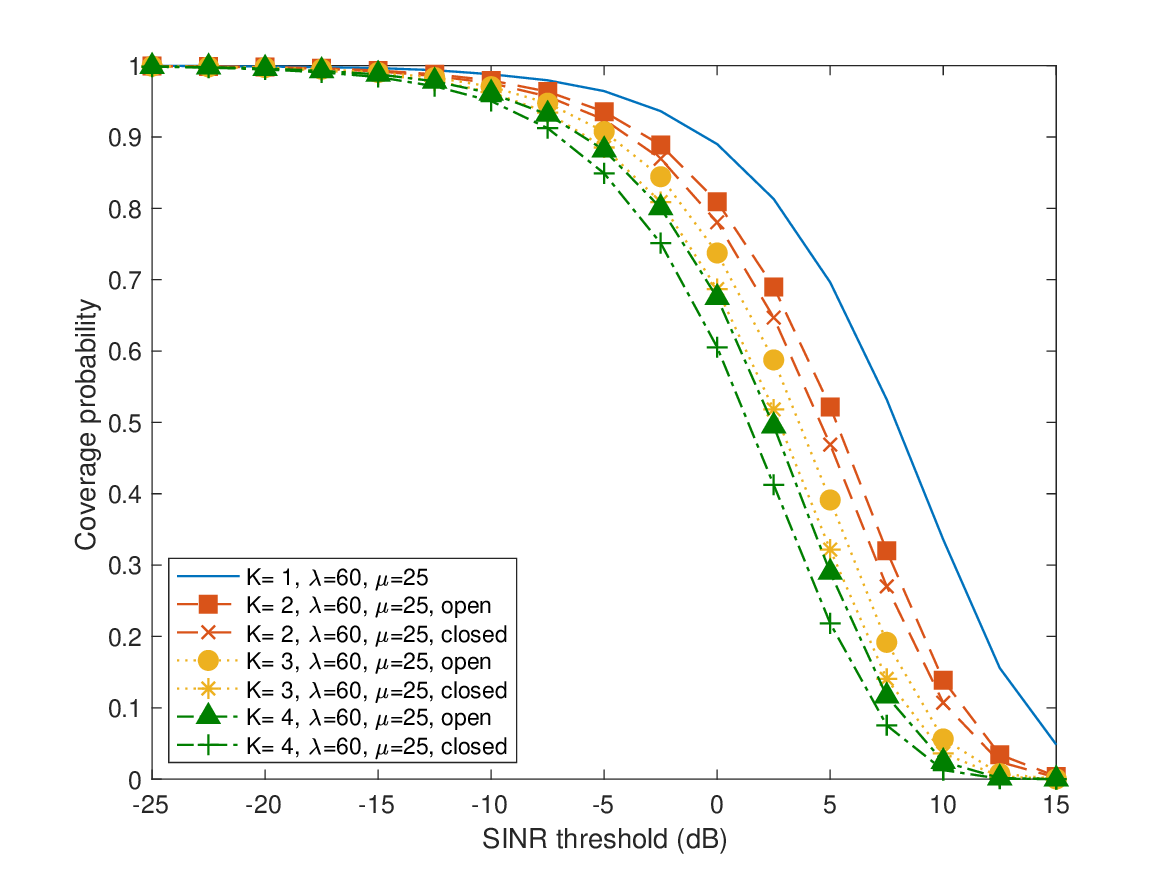}
		\caption{Illustration of the coverage probability. The satellite altitude is  $550$ km. $g=20$ dB.}
		\label{fig:siropenvsclosedtotalantenna23db}
	\end{figure}
	

\section{Discussion}
\subsection{Compare to an Upcoming Constellation}
In this section, we employ the proposed Cox point process to model upcoming Starlink and OneWeb constellations. We compare the coverage probabilities between the Cox model and these future heterogeneous constellations, aiming to demonstrate the efficacy of our proposed model. Our objective is to illustrate how the proposed Cox model can accurately represent complex forthcoming LEO satellite constellations and predict the typical performance behavior.

Initially, we utilize online specifications for the Starlink (frequency reuse factor: $8$) and OneWeb constellations, presented in Table \ref{Table:STLK}. To represent the orbital distributions and satellites of these constellations using Cox point processes, we enhance a moment matching method in \cite{choi2023Cox} which aligns the number of orbits or the number of satellites of the Cox point process to ensure that the Cox model approximate the local distribution of LEO satellites in those forthcoming constellations. 

For both practical constellations and the proposed Cox point process, network performance remains independent of user longitudes. For the upcoming constellations, the visibility of satellites varies based on the user latitudes.  On the other hand, the proposed Cox point process is rotation invariant and therefore the network performance is independence from the user latitudes. This paper focuses on two distinct user distributions: one at the equator and another at a latitude of $30$ degrees. Additionally, due to the substantially higher number of Starlink LEO satellites compared to OneWeb LEO satellites, we consider a frequency reuse factor of $8$. We assume that $1/8$ Starlink LEO satellites, share the same frequency resources with $648$ OneWeb satellites and they interfere each other.

The average number of Starlink satellites visible from users at latitude $30$ degrees is $60$, while for the OneWeb constellation, the average number of visible satellites is $45$. At the user latitude of $0$ degree, the average number of visible Starlink satellites is $38$ whereas the average number of visibile OneWeb satellites is $34$. Based on these empirical mean values, we determine ${\lambda}_{i=1,2}$ and ${\mu}_{i=1,2}$ such that the Cox point processes representing the constellations have the same average numbers of visible satellites. This moment matching method ensures that the Cox model locally approximates the forthcoming constellations by having the same numbers of visible satellites.

Figure \ref{fig:starlinkoneweb} illustrates the two-tier heterogeneous LEO satellite network with Starlink and OneWeb, while Figures \ref{fig:section6lat0} and \ref{fig:section6lat30} depict the coverage probability of the actual deployment and the proposed Cox point process for various user latitudes. In both scenarios, we confirm that the proposed Cox model reproduces the coverage probability of the closed access probability for the typical user in the Starlink constellation. A difference less than $1$ dB is observed for both figures and this difference stems from the geometric difference between the isotropic Cox point process and the non-isotropic Starlink constellation. Please note that the minimum distance between satellites in orbit might also contribute to a slightly improved coverage probability in the Starlink scenario, as discussed in \cite{10058140}. A more detailed analysis on the comparison between the Cox point process and forthcoming future satellite constellations is left for future work.  

\subsection{Isotropic Constellation}
The surge in interest surrounding LEO satellites initially aimed to serve diverse network users may lead to isotropic orbits in the future. Recent successful demonstrations of LEO satellite network applicability have prompted a growing number of companies and governments to express interest in deploying their own LEO satellites. For instance, Space-X, plans to implement a second generation with a different geometric configuration (altitudes, inclinations, longitudes) to prevent satellite collisions and enhance service to subscribers, facing with increasing congestion of LEO satellite orbits.  With multiple entities focusing on various applications, this trend is expected to persist and intensify, resulting in crowded orbits. This congestion may lead to a completely random orbit, resembling the proposed Cox point process.

An intriguing interpretation of Cox-distributed orbits is that the proposed Cox model can be interpreted as a maximum entropy model. In the proof of isotropy for the Cox point process \cite{choi2023Cox}, the proposed orbit process is expressed as a factor of a uniformly distributed random point on a sphere $\mathbb{S}$. Provided that the uniformly distributed point process has the maximum entropy among all point processes on $\mathbb{S}$, the proposed orbit process, by extension, would have the maximum entropy among all orbit processes on $\mathbb{S}$. Consequently, the isotropic model may be a suitable representation for future constellations featuring numerous operators and satellites, particularly when the orbits of constellations are deployed randomly and independently. A more thorough investigation into this aspect is left for future work.

	\begin{figure}
		\centering
		\includegraphics[width=1\linewidth]{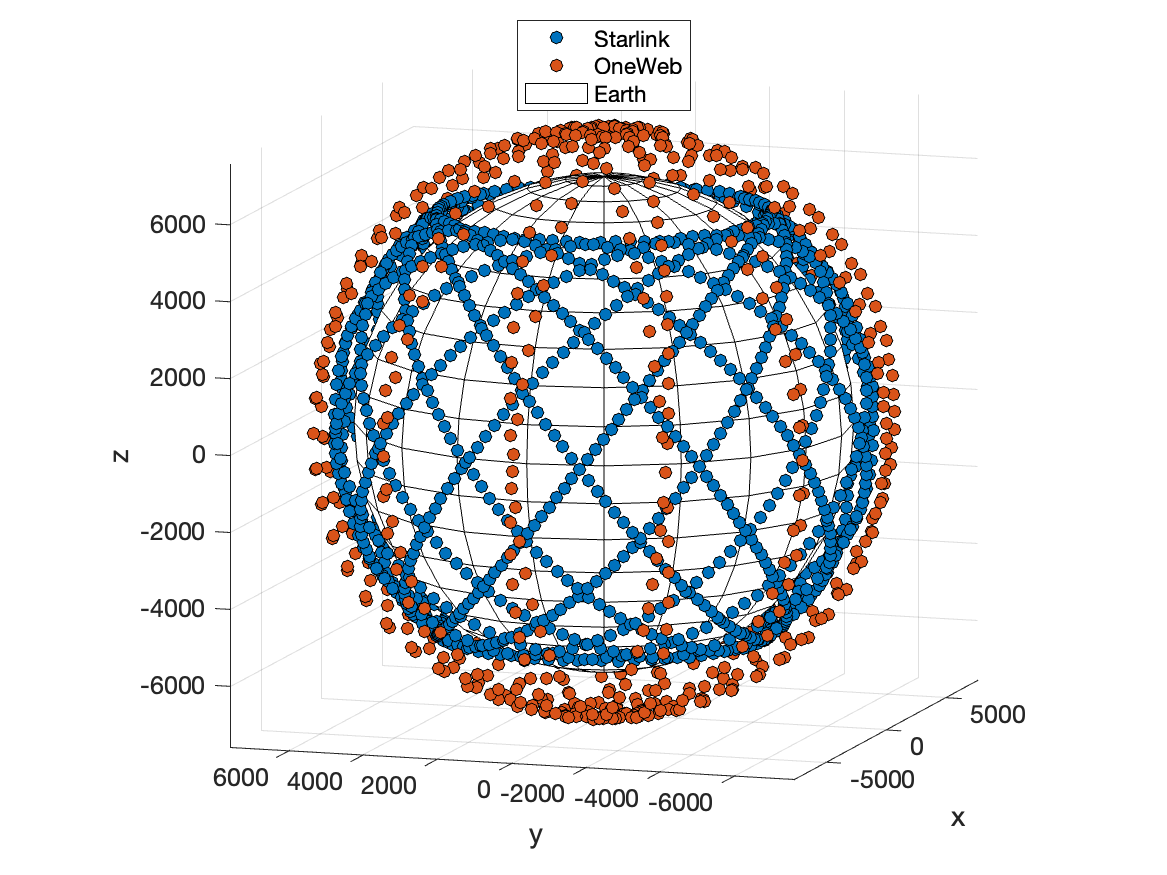}
		\caption{Starlink 2A group 1 and OneWeb constellations.}
		\label{fig:starlinkoneweb}
	\end{figure}

	\begin{figure}
		\centering
		\includegraphics[width=1\linewidth]{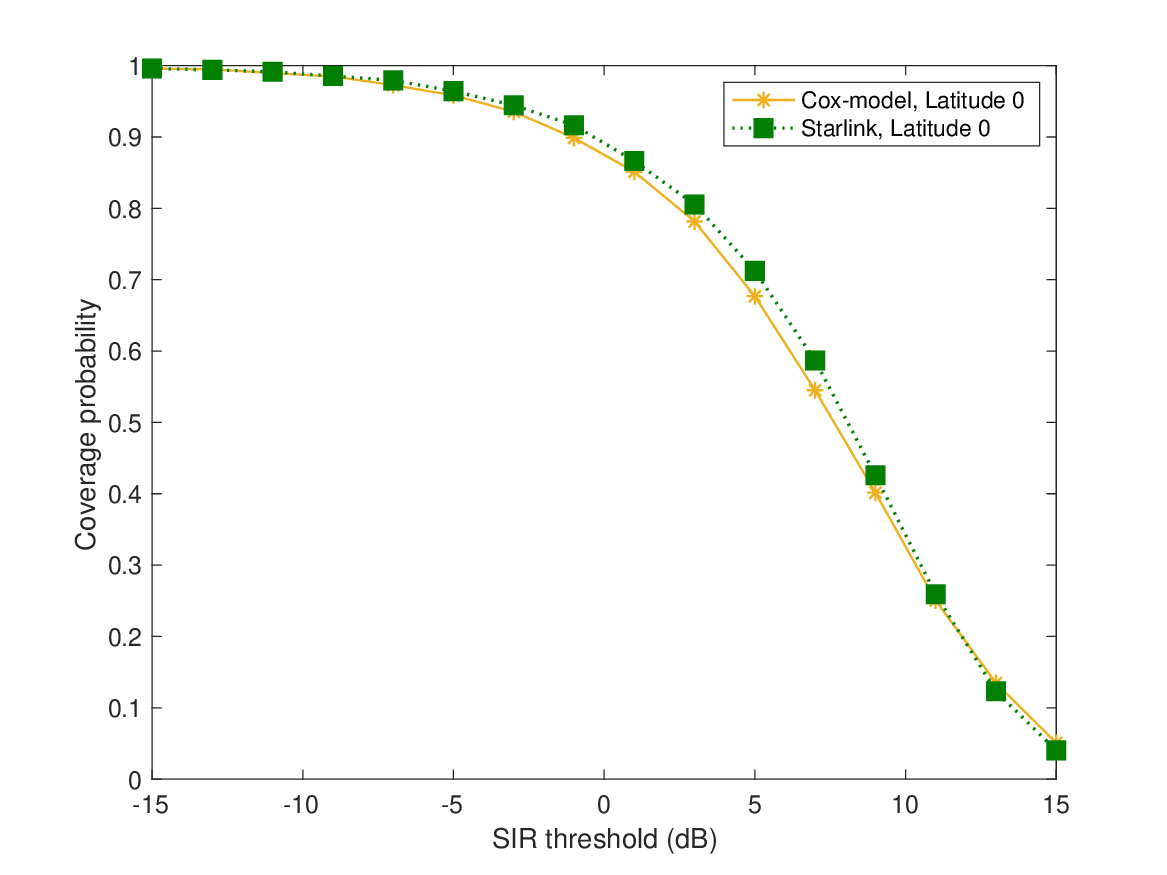}
		\caption{The closed access coverage probabilities for the Starlink deployment and the Cox model. The users are at a latitude of 0 degrees. }
		\label{fig:section6lat0}
	\end{figure}
	
	\begin{figure}
		\centering
		\includegraphics[width=1\linewidth]{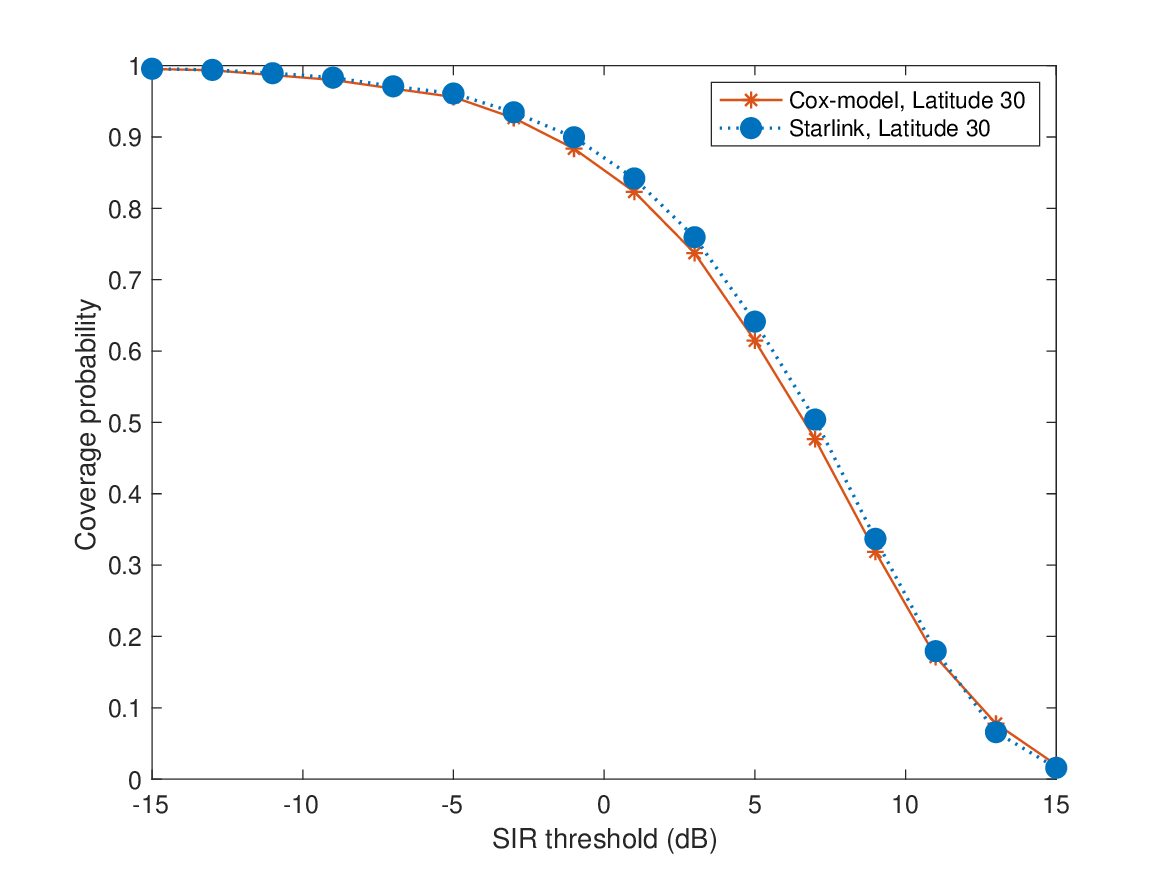}
		\caption{The closed access coverage probabilities for the Starlink and the Cox model. The users are located at a latitude of 30 degrees.}
		\label{fig:section6lat30}
	\end{figure}

		\begin{table}
		\caption{Starlink and OneWeb Constellation}\label{Table:STLK}
		\centering 
		\begin{tabular}{|c|c|c|c|c|c|}
			\hline
			Name & Altitude & Incl. & Count  &  Sat.  & Total \\
			\hline
			Starlink 2A group $1$ & $530$ km & $43\degree$  & $28$ & $120$  &$3360 $\\\hline
			Starlink 2A group $2$ & 525 km & $53\degree$ & $28$ & $120$  &$3360$ \\\hline
			Starlink 2A group $3$ & 535 km & $33\degree$ & $28$ & $120$  &$3360$ \\\hline
			OneWeb & 1200 km & $86.4\degree$ & $12$ & $54$ & $648$ \\\hline
		\end{tabular}
	\end{table}
\section{Conclusion}
This research paper utilizes Cox point processes to model and analyze LEO satellite constellations deployed at various altitudes and quantities. The paper explores two user access technologies: close access and open access. In the closed access scenario, users can only communicate with LEO satellites of the same type, and local interference may arise from LEO satellites of different types. The paper calculates the interference and SINR coverage probability for the typical user in each type, revealing that as the number of satellite types increases, the coverage probability experiences a significant decline due to local interference. To address this issue, the paper proposes an open access scenario where users can communicate with any type of LEO satellite. The association probability for the typical user is calculated, and its coverage probability is derived. The paper concludes that the average SINR in the open access scenario is nearly double that of closed access, and this substantial SINR improvement applies to all percentile users. The analysis emphasizes the detrimental impact of local interference in closed access downlink communication of LEO satellite networks with multiple constellations while showcasing the advantages of an open access scenario. The findings of this paper offer valuable insights into the design and optimization of heterogeneous LEO satellite networks with multiple coexisting operators.

As for future work, the paper suggests several extensions. These include utilizing another nonhomogeneous orbit process to model practical satellite orbit systems, developing an interference management technique using the Laplace transform of interference, and designing a load-aware advanced open access system with a switching protocol that allows users to choose between closed and open access schemes based on the proximity of nearby ground downlink communication terminals. Leveraging the proposed framework for a heterogeneous LEO satellite network, one can study the medium access control or transport layer aspects of such a network including inter-satellite links, or traffic offloading.

%
%
%
%
%
%
%
%

%
%

	\section*{acknowledgement}
	This work was supported by the National Research Foundation of Korea(NRF) grant funded by the Korea Government(Ministry of Science and ICT)(No. NRF-2021R1F1A1059666 and No. RS 2023-00247692). 

	\bibliographystyle{IEEEtran}
	\bibliography{ref}

\end{document}